\documentclass[iop]{emulateapj}

\usepackage{times}
\usepackage{graphicx}
\usepackage{amsmath}
\usepackage{float}
\usepackage{ifthen}
\usepackage{times}
\usepackage{natbib}
\usepackage{rotating}
\usepackage{units}
\newcommand{\fircolour}{$f_{60}/f_{100}$\,}

\newcommand{\ci}{\,{\rm [C\,{\sc i}]($^3P_1$$\rightarrow^3$$P_0$)}\,}
\newcommand{\cii}{{\rm [C\,{\sc ii}]\,}}%($^3P_1\rightarrow ^3P_0$)}\,}
\newcommand{\coiv}{\,{\rm CO\,{\it J}($4$$\rightarrow$$3$)}\,}

\newcommand{\coi}{\,{\rm CO\,{\it J}($1$$\rightarrow$$0$)}\,}
\newcommand{\coii}{\,{\rm CO\,{\it J}($2$$\rightarrow$$1$)}\,}
\newcommand{\coix}{\,{\rm CO\,{\it J}($9$$\rightarrow$$8$)}\,}
\newcommand{\cox}{\,{\rm CO\,{\it J}($10$$\rightarrow$$9$)}\,}
\newcommand{\coiii}{\,{\rm CO\,{\it J}($3$$\rightarrow$$2$)}\,}
\newcommand{\hcni}{\,{\rm HCN\,{\it J}($1$$\rightarrow$$0$)}\,}

\newcommand{\hcniv}{\,{\rm HCN\,{\it J}($4$$\rightarrow$$3$)}\,}

\begin{document}

\shorttitle{Molecular and atomic line surveys of galaxies I}
\shortauthors{Geach \& Papadopoulos}

\title{Molecular and atomic line surveys of galaxies I:\\ The dense, star-forming phase as a beacon }

\author{James~E.~Geach\altaffilmark{1,2} and Padelis \ P.\
Papadopoulos\altaffilmark{3}}

\altaffiltext{1}{Banting Fellow}

\altaffiltext{2}{Department of Physics, McGill University,
3600 rue University, Montr\'eal, Qu\'ebec, H3A 2T8, Canada.
jimgeach@physics.mcgill.ca}

\altaffiltext{3}{Max Planck Institute for Radioastronomy, Auf dem H\"ugel 69,
D--53121 Bonn, Germany. padelis@mpifr-bonn.mpg.de}

\begin{abstract}We predict the space density of molecular gas reservoirs in
the Universe, and place a lower limit on the number counts of carbon monoxide
(CO), hydrogen cyanide (HCN) molecular and \cii atomic emission lines in blind
redshift surveys in the submillimeter--centimeter spectral regime. Our model
uses:\ (a) recently available HCN Spectral Line Energy Distributions (SLEDs)
of local Luminous Infrared Galaxies (LIRGs, $L_{\rm IR}>10^{11}L_\odot$), (b)\
a value for $\epsilon_{\star}$=$\rm SFR$/$M_{\rm dense}(\rm H_2)$ provided by
new developments in the study of star formation feedback on the interstellar
medium and (c)\ a model for the evolution of the infrared luminosity density.
Minimal `emergent' CO SLEDs from the dense gas reservoirs expected in all
star-forming systems in the Universe are then computed from the HCN SLEDs
since warm, HCN-bright gas will necessarily be CO-bright, with the dense
star-forming gas phase setting an obvious minimum to the total molecular gas
mass of any star-forming galaxy. We include \cii as the most important of the
far-infrared cooling lines. Optimal blind surveys with the Atacama Large
Millimeter Array (ALMA) could potentially detect very distant
($z\sim10$--$12$) \cii emitters in the $\geq$ULIRG galaxy class at a rate of
$\sim$0.1--1 per hour (although this prediction is strongly dependent on the
star formation and enrichment history at this early epoch), whereas the
(high-frequency) Square Kilometer Array (SKA) will be capable of blindly
detecting $z>3$ low-{\it J} CO emitters at a rate of $\sim$40--70 per hour.
The \cii line holds special promise for the detection of metal-poor systems
with extensive reservoirs of CO-dark molecular gas where detection rates with
ALMA can reach up to 2--7 per hour in Bands 4--6.\end{abstract}

\section{Introduction}

Since the first detections of the $J$$=$$1$$\rightarrow$$0$ rotational transition of
$^{12}$CO and some of its isotopologues in Galactic molecular clouds
($^{13}$CO, C$^{18}$O) (Wilson et al.\ 1970; Penzias et al.\ 1971, 1972), and
in galactic nuclei (Rickard et al.\ 1975) there have been many studies of CO
line emission in galaxies using single dish radio telescopes and
interferometer arrays (for reviews see Young \& Scoville\ 1991 and Solomon \&
Vanden Bout\ 2005). Multi-{\it J} CO line ratio surveys are now routinely used
to assess the state of the molecular gas in galaxies (e.g.\ Braine \& Combes\
1992; Aalto et al.\ 1995; Papadopoulos \& Seaquist\ 1998; Mauersberger et al.\
1999; Nieten et al.\ 1999; Yao et al.\ 2003; Mao et al.\ 2011) over the
density regime where most of its mass resides ($n\sim 10^{2-3}\,{\rm
cm}^{-3}$). The fainter molecular line emission from heavy-rotor molecules
such as HCN have also become a standard tool for assesing the state and the
mass of the denser ($>$$10^4\,{\rm cm}^{-3}$) gas phase where stars actually
form in Giant Molecular Clouds (GMCs, e.g.\ Nguyen-Q-Rieu et al.\ 1989;
Solomon et al.\ 1992a; Paglione et al.\ 1995, 1997; Jackson et al.\ 1995). The
role of the latter phase as the direct fuel of star formation in individual
GMCs, quiescent disks and merger-driven spectacular starbursts in the local
and distant Universe is now well established over an astounding 7--8 orders of
magnitude (Gao \& Solomon\ 2004; Wu et al.\ 2005; Juneau et al.\ 2009; Wang et
al.\ 2011).

 In the past decade, numerous high-{\it z} detections have revealed the
fundamental role of molecular lines in assessing the state and mass of the
molecular gas, and the dynamical mass of heavily dust-enshrouded galaxies in
the early Universe (Solomon \& Vanden Bout 2005\ and references therein), and
in some cases have provided remarkable insights into the properties of the
molecular interstellar medium (ISM) in early galaxies (see Danielson et al.\
2010 for a recent example of a well sampled CO spectral line energy
distribution (SLED) in a $z=2.3$ gravitationally lensed galaxy). This
exploration began with the first detection of \coiii\ line emission in the
strongly-lensed distant dust-enshrouded galaxy IRAS\,10214+4724 at $z\sim 2.3$
(Brown \& Vanden Bout\ 1991; Solomon et al.\ 1992b). It continued with the
detection of CO transitions in distant submm-bright galaxies (SMGs, $L_{\rm
IR}>10^{12}L_\odot$) (Frayer et al.\ 1998, 1999; Greve et al.\ 2005) and is
increasingly encompassing less extreme, but still massive systems such as
Lyman Break galaxies (Baker et al.\ 2004), optical/near-infrared selected
galaxies at $z\sim1.5$ (Dannerbauer et al.\ 2009; Daddi et al.\ 2010) and
fortuitously lensed systems (Danielson et al.\ 2011; Lupu et al.\ 2011).
Several spectacular CO line detections have also been obtained also for other
high-redshift systems such as radio galaxies (e.g.\ De Breuck et al.\ 2005)
and QSOs out to $z\sim 6.4$. This epoch is close to the era of reionization --
the final frontier of galaxy evolution studies -- revealing the gas-rich hosts
to rapid galaxy growth at these early times (Walter et al.\ 2003, 2004; Weiss
et al.\ 2007).

These discoveries and advancements, made possible as sensitivities of
millimeter/submillimeter interferometer arrays improved in the last decade,
still yield only a glimpse of what will be a new era where molecular and
atomic (e.g.\ \cii\,$\lambda158$, \ci) ISM lines will replace nebular
optical/near-infrared (OIR) lines as the main tool of choice for discerning
galaxy formation and evolution across the full span of cosmic time pertinent
to galaxy growth, from the end of the reionisation epoch ($z>7$) to the
present (Walter \& Carilli\ 2008).

Direct `blind' searches of gas-rich galaxies using submm--cm wave molecular
and atomic lines are the only tool that can:\ (i) uniformly select galaxies
according to their molecular gas content rather than their SFR and the star
formation efficiency (a bias that has so far -- necessarily -- affected all
high-$z$ gas studies),\ (ii) immediately provides redshifts and eventually
dynamical mass information,\ (iii) holds the promise of discovering large
outliers of the local $\Sigma _{\rm SFR}$-$\Sigma_{{\rm H}_2}$
(Schmidt--Kennicutt) relations (Kennicutt\ 1998), with large reservoirs of
molecular gas but low levels of SFR (Papadopoulos \& Pelupessy\ 2010), and\
(iv) can possibly determine the star formation `mode' (starburst/merger-driven
versus quiescent/disk-like), in a uniform and extinction-free manner (Paper
II, Papadopoulos \& Geach\ 2012).

Well-sampled CO spectral line energy distributions, and their robust
normalization by some observable galaxy property are necessary for predicting
the emergent CO line luminosities in star-forming systems. The lack of these
two key ingredients translates to major uncertainties for the source counts
predicted for blank-field cm/mm/submm molecular line surveys (Combes et al.\
1999; Blain et al.\ 2000; Carilli \& Blain\ 2002), as well as the frequency
and flux range where such surveys become optimal (Blain et al.\ 2000). The
dense gas phase ($n({\rm H}_2)$$>$10$^{4}$\,cm$^{-3}$) and the linear relation
of its mass to the SFR in individual GMCs ($L_{\rm IR}$$\sim
$10$^{4.5}$\,$L_{\odot}$), ultraluminous infrared galaxies (ULIRGs, $L_{\rm
IR}$$\sim $$10^{12}$\,$L_{\odot}$) and high-redshift extreme starbursts
(HLIRGs, $L_{\rm IR}$$\sim $$10^{13}$\,$L_{\odot}$) makes it an obvious
benchmark for computing minimum emergent molecular line luminosities. Indeed
only this gas phase yields physically meaningful estimates of the so-called
star formation efficiency (and its equivalent interpretation in terms of gas
consumption timescales) while the HCN-deduced (and thus well-excited CO SLEDs)
contain minimal uncertainties up to high-{\it J} rotational transitions of CO.
 The dense, HCN-bright, molecular gas phase in galaxies is thus an obvious
ingredient of any theoretical models for blank-field cm/mm/submm molecular
line surveys.

\section{Objectives of this work}

  In this work we compute the number counts of the star-forming molecular gas
  reservoirs in the Universe using:\ (i) hydrogen cyanide (HCN) SLEDs of local
  luminous infrared galaxies (LIRGs, $L_{\rm IR}$$\sim
  $$10^{11}$\,$L_{\odot}$), and (ii) an $\epsilon_{\star}$=$\rm SFR$/$M_{\rm
  dense}$(H$_2$) value provided by recent studies of star formation feedback
  on the interstellar medium (ISM). The latter is crucial for relating the
  cosmic SFR({\it z}) to the dense gas mass necessary to fuel it. Minimal
  emergent CO SLEDs up to \cox, normalized by the galaxy SFRs, can then be
  computed even from partial low-{\it J} HCN SLEDs ($J_{\rm up}\leq4$) since
  warm HCN-bright gas will also be CO-bright ($n_{\rm crit}$(HCN)/$n_{\rm
  crit}$(CO)$>$100), while the dense HCN-bright star-forming gas mass sets an
  obvious minimum to the total molecular gas mass of star-forming systems.

These SFR-normalized minimal CO SLEDs can then be used as inputs to various
galaxy-evolution models to yield minimum source counts in blank-field
cm/mm/submm molecular line surveys of star-forming systems. We use an
empirically based phenomenological model for the evolution of the bolometric
(IR) luminosity function that accurately re-produces the observed number
counts of galaxies in several infrared and sub-millimeter surveys ({\it
Spitzer}, {\it Herschel} and SCUBA) to predict the number counts of molecular
line-emitting galaxies seen by the Atacama Large Millimeter Array (ALMA), the
Jansky Very Large Array (JVLA) and the Square Kilometer Array (SKA) and its
pathfinders. Using the shape of the integral numbers counts as a guide, we
suggest the strategy for an optimal blind redshift survey (e.g.\ Blain et al.\
2000; Carilli \& Blain\ 2002) that could detect gas-rich galaxies across the
full history of galaxy evolution. Throughout we assume a $\Lambda$CDM
cosmological model, with $\Omega_{\rm m}=0.3$, $\Omega_\Lambda=0.7$ and
$H_0=70$\,km\,s$^{-1}$\,Mpc$^{-1}$.

\begin{figure}[t]
\centerline{\includegraphics[width=0.5\textwidth,angle=-90]{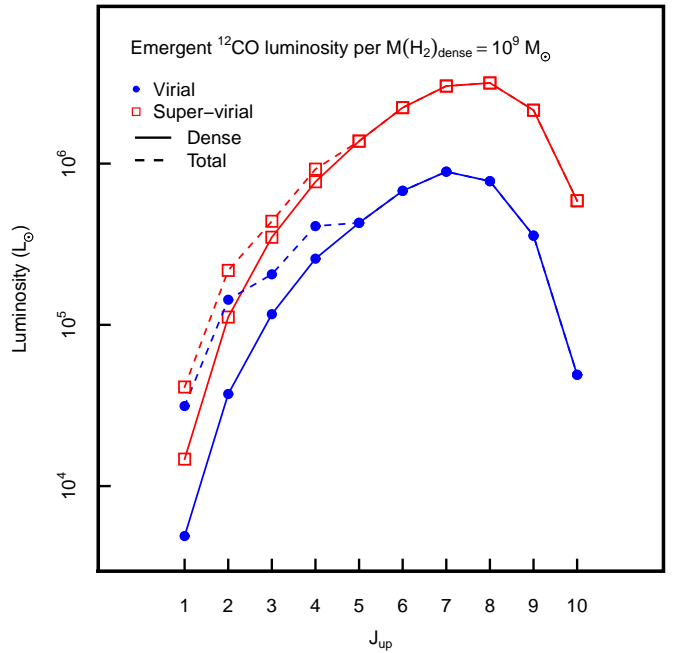}}
\caption{Emergent $^{12}$CO spectral line energy distributions for a gas mass
of $M_{\rm dense} = 10^9M_\odot$, for our virial ($X_{\rm CO}=9\,M_\odot\,{\rm
(K\,km\,s ^{-1}\,pc^2)^{-1}}$) and super-virial ($X_{\rm CO}=3\,M_\odot\,{\rm
(K\,km\,s ^{-1}\,pc^2)^{-1}}$) models. The solid line shows just the
emergent flux for the dense phase, and the dashed line shows the total
emission when the quiescent (i.e.\ cold SLED, $X_{\rm CO}=5\,M_\odot\,{\rm
(K\,km\,s ^{-1}\,pc^2)^{-1}}$)  phase is included (see
\S3.2). This is the most conservative estimate for the molecular line
emission from a star-forming galaxy, with the range in CO luminosity
given by the two models
reflecting the possible range in emission that would arise from
different dynamic configurations of the cold ISM in active galaxies
(\S3.1).} \label{fig:sled} \end{figure}

\begin{figure}
\centerline{\includegraphics[height=0.52\textwidth,angle=-90]{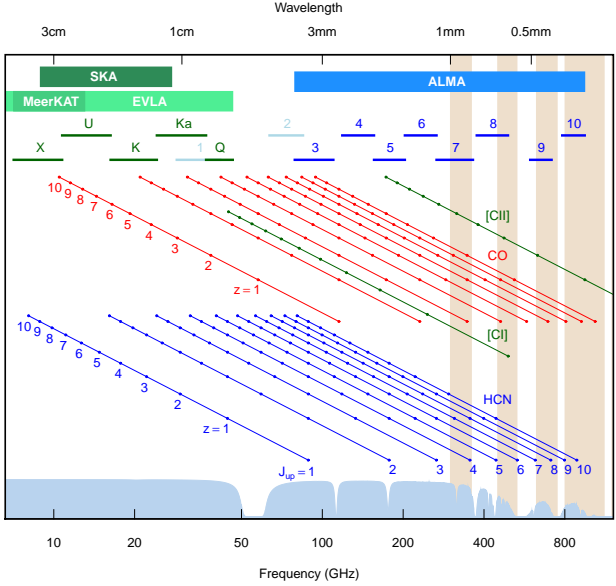}}
\caption{Schematic view of the visibility of the $J_{\rm up}\leq 10$
transitions of CO and HCN to $z\leq10$, and the \ci and \cii fine
structure lines, indicating the band coverage offered
by the SKA high-frequency component and its pathfinder MeerKAT, as well as the
radio bands {\it X, U, K, Ka, Q} and ALMA Bands 3--10. SKA, MeerKAT
and JVLA
offers the ability to detect low-{\it J} CO and HCN emitting galaxies out to
very high redshifts, whereas ALMA covers a broader range of transitions up to
high-{\it J} over a large span of cosmic time, but can access low-{\it J
lines} only out to $z\lesssim3$. The vertical bars show the bandpasses
of ground-based instrumentation that could access the FIR lines in the
sub-mm regime (e.g.\ the Redshift and Early Universe Spectrometer
[ZEUS], see Stacey et al.\ 2007).} \label{fig:schem} \end{figure}

 \begin{figure*}
\centerline{\includegraphics[height=0.99\textwidth,angle=-90]{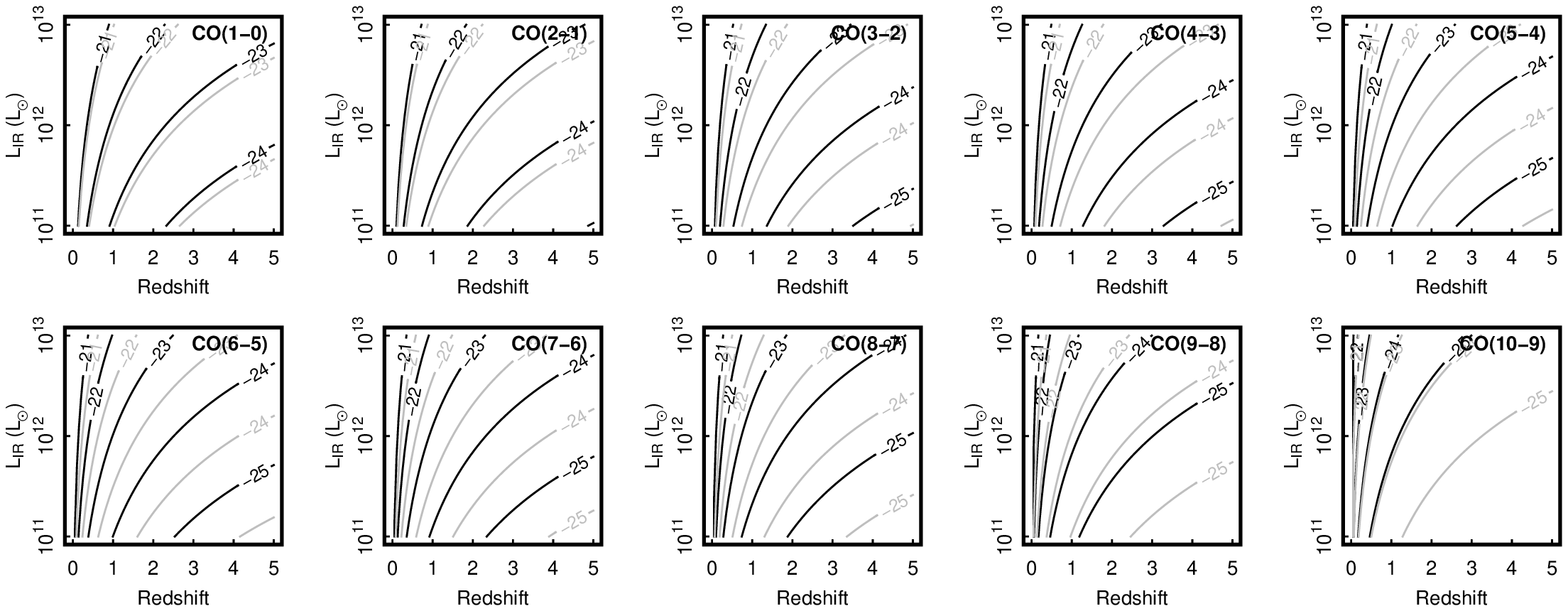}}
\caption{Predicted CO line fluxes for luminous infrared galaxies as a
function of redshift (shown as contours of constant line flux with units of
$\log(F/{\rm W\,m^{-2}})$ labelled) from our emergent model. Bold lines show
the `virial' model prediction, and grey lines show the equivalent
`super-virial' model. Given the sensitivities of submm/mm/cm
instrumentation, this model can be used as a conservative estimate of
the time required to detect a CO emission line from star-forming
(infrared-bright) galaxies
across cosmic time. This will be valuable in follow-up spectroscopy of
submm continuum surveys to be conducted with, e.g.\ JCMT, LMT \& CCAT that
will track LIRG--ULIRG level activity out to $z\sim5$.}
\label{fig:grid} \end{figure*}

\section{The molecular ISM model: a minimalist approach }

\subsection{The dense, star-forming gas phase}

Past studies used rather poorly constrained models of the molecular phase ISM
and its CO line excitation range to derive the emergent line luminosities
(e.g.\ Combes et al. 1999; Blain et al. 2000). Here we make use of available
HCN {$J_{\rm up}\leq4$ SLEDs of LIRGs (Papadopoulos et al.\ 2007; Krips et
al.\ 2008; Juneau et al.\ 2009) to safely extrapolate the corresponding
emergent CO SLEDs from \coi to \cox for the dense star-forming gas in
galaxies. This is possible since even a modestly excited global \hcniv line
emission (e.g.\ HCN\ $(4\rightarrow3)/(1\rightarrow0)=0.30$) implies gas
densities $n({\rm H}_2)\sim 3\times 10^4-10^5$\,cm$^{-3}$ (e.g. Papadopoulos
et al.\ 2007), similar to the critical density of \coix ($n_{\rm
cr}$=$2.2\times 10^5$$\rm cm^{-3}$). Furthermore, unlike past studies, we use
our Large Velocity Gradient (LVG) radiative transfer code to constrain the
properties of the dense gas phase using observed global HCN line ratios in
LIRGs {\it and} $K_{\rm vir}$$\sim $1 where

	\begin{align}
\nonumber 	K_{\rm vir}({\rm HCN}) & =\frac{\left({\rm d}V/{\rm d}R\right)}{\left({\rm d}V/{\rm d}R\right)_{\rm vir}} \\ 
	& \sim
	1.54\frac{{\rm [HCN/H_2]}}{\sqrt{\alpha}\Lambda
	  _{\rm HCN}}\left(\frac{n({\rm H_2})}{10^3\, {\rm cm}^{-3}}\right)^{-1/2},
	\end{align}

\noindent parametrizes the average dynamic state of the corresponding gas,
with $K_{\rm vir}$$\sim $1 corresponding to virial gas motions (typical for
dense HCN-bright star-forming cores in the Galaxy), and $K_{\rm vir}$$\gg$1
corresponding to unbound gas (typically found for low-{\it J} CO line emission
in LIRGs). The parameter $\alpha\sim 0.55$--$2.4$ depends on the average cloud
density profile (see Bryant \& Scoville 1996), here we choose $\alpha =1.5$.
The quantity $\Lambda_{\rm HCN}={\rm [HCN/H_2]}/({\rm d}V/{\rm d}R)$ is one of
the three parameters defining the grid of a typical LVG code (the other two
being $T_{\rm kin}$ and $n({\rm H}_2)$). Furthermore we adopt $T_{\rm
kin}$$\sim $$T_{\rm dust}$, which is certainly a good approximation for the
dense gas phase where strong gas-dust thermal coupling sets in. Abundances of
$\rm [HCN/H_2]$=$2\times 10^{-8}$ and $\rm [CO/H_2]$=10$^{-4}$ are adopted for
our estimates of $K_{\rm vir}$ for HCN and CO LVG solutions.

Multi-{\it J} HCN line surveys of LIRGs find average $r_{J+1,J}({\rm
HCN})={\rm HCN}\,(J+1\rightarrow J)/(1\rightarrow 0)$ brightness temperature
ratios of $r_{21}{\rm (HCN)}\sim 0.65$ and $r_{32}{\rm (HCN)}\sim 0.45$ (Krips
et al.\ 2008; Juneau et al.\ 2009). These can be higher still (up to $\sim$1)
for extreme mergers whose ISM is dominated by very dense gas (Greve et al.
2009), but the aforementioned ratios are good representatives of the average
values observed in LIRGs and we use them as constraints on the properties of
the HCN-bright star-forming dense gas phase. The LVG grid is run for
temperatures of $T_{\rm kin}=(30-80)$\,K fully encompassing the typical range
expected for star-forming gas and its concomitant dust. We find $n({\rm
H}_2)=3\times 10^4$\,cm$^{-3}$, $T_{\rm kin}=40$\,K, and $K_{\rm vir}\sim 1$
as well as $n({\rm H}_2)=10^5$\,cm$^{-3}$, $T_{\rm kin}=(30$--$35)$\,K and
$K_{\rm vir}\sim 8$ as typical solutions. The virial solution is the most
likely one for a dense star-forming gas phase in LIRGs (in current
turbulence-regulated star-forming models of galaxies self-gravity overcomes
turbulent gas motions at such high densities, Krumholz \& McKee 2005). We
nevertheless consider also the solution with $K_{\rm vir}$$\sim $8,
corresponding to gravitationally unbound gas motions; this `super-virial'
solution is possibly pertinent to the high density gas found in some extreme
starbursts at high redshifts (Swinbank et al.\ 2011). Any additional
excitation mechanisms such as hard X-rays from an active galactic nuclei (AGN;
Meijerink \& Spaans 2005), and/or high cosmic ray energy densities due to the
high supernovae rate densities expected in compact ULIRGs (Papadopoulos 2010)
can only raise the dense gas temperatures leading to even more excited HCN and
CO SLEDs.

 Finally we note that for simplicity we omit the enhanced cosmic microwave
background (CMB) as it will have negligible effects on the dense and warm
star-forming gas phase in galaxies, even at high redshifts. For a detailed
discussion on this see Papadopoulos et al.\ 2000 (section 4.2), but this can
be briefly demonstrated by assuming $T_{\rm kin}=40$\,K and $T_{\rm kin}\sim
T_{\rm dust}$, typical for the dense, HCN-bright, star-forming gas in LIRGs.
The thermodynamically equivalent $T(z)=T_{\rm dust}(z)\sim T_{\rm kin}(z)$ of
that phase at $z=5$ would then be: $T(z) = {(T(0)^6-T_{\rm CMB}(0)^6+T_{\rm
CMB}(z)^6)}^{1/6} \approx 40$\,K, i.e.\ it remains essentially identical (a
dust emissivity law of $a=2$ is assumed). The impact can be more significant
for CO line ratios of a cold and quiescent gas phase (for a $T_{\rm
kin}(0)=15$\,K, $T_{\rm kin}(z=5)=17.56$\,K) where these ratios can be boosted
somewhat by the enhanced CMB (see also Table\ 2 in Papadopoulos et al.\ 2000),
while the line/CMB contrast for the cold, quiescent H$_2$ phase diminishes. In
that regard our assumed local CO line ratios for that phase (see following
section) remain conservative and in the spirit of our minimalist approach.

Here we must note that any such global LVG solutions reflect, at best, a rough
average of the gas conditions responsible for the observed global line ratios
considered. Indeed, as is the case for the so-called $X_{\rm CO}$, $X_{\rm
HCN}$, etc. factors that convert the CO and HCN $J=1\rightarrow 0$ line
luminosities to mass for the total and the dense molecular gas reservoirs
respectively, these LVG solutions represent effective ensemble averages for
large collections of molecular clouds, and thus appropriate for unresolved
studies of high-$z$ galaxies. In Table\,1 we list the LVG-derived line ratios
that correspond to the two types of LVG solutions found for the dense gas.
These have very similar HCN line ratios up to $J_{\rm up}=5$ (we do not
extrapolate above this transition as data for only up to \hcniv exist in the
local Universe), and CO up to $J_{\rm up}=4$. Above $J_{\rm up}=4$ we give the
range of acceptable values yielded by the LVG model. The corresponding $X_{\rm
CO, dense}$ factors are: $\sim$9\,$
M_{\odot}$\,(K\,km\,s$^{-1}$\,pc$^{2}$)$^{-1}$ and $\sim $3\,$
M_{\odot}$\,(K\,km\,s$^{-1}$\,pc$^{2}$)$^{-1}$ for the virial and the
super-virial LVG solution respectively, and along with the computed line
ratios (Table 1) are used to obtain the range of CO SLEDs for the dense gas
shown in Figure\,1.

The computed CO SLEDs are minimal in two ways, namely:

\begin{enumerate}

\item Only a fraction of the total molecular gas mass in galaxies belongs to
this dense star-forming gas phase contributing to such SLEDs at any given
moment of their evolution.

\item The observed global HCN line ratios used for their derivation
inadvertently include some non-star-forming, less dense gas diluting the HCN
ratios from those typical for the star-forming phase alone.

\end{enumerate}

The HCN SLEDs are considered only up to $J_{\rm up}=5$ since:\ (a) available
HCN line data for LIRGs in the local Universe exist only up to $J_{\rm up} =4$
(e.g.\ Papadopoulos\ 2007), and\ (b) only HCN lines up to $J_{\rm up}=3$ were
used in our LVG model (thus interpolation beyond $J_{\rm up}=5$ is not safe).
We also note that the normalization of the HCN SLED, namely the $X_{\rm
HCN}$=$M_{\rm dense}({\rm H_2})/L' _{\rm HCN}(1\rightarrow0)$ factor is not
identical to that of CO SLEDs since the \hcni is not fully thermalized (unlike
\coi) even at the high densities of our LVG solutions. The latter yield
$X_{\rm HCN}$=\{9, 27\}\,$M_\odot$\,(K\,km\,s$^{-1}$\,pc$^{2}$)$^{-1}$ for the
unbound (i.e. super-virial) and the virial solution respectively.

The total CO SLED of a star-forming galaxy will also include non-star-forming
gas in a much more tenuous phase. This typically contains the bulk of the
molecular gas in individual GMCs, and the molecular gas reservoirs in local
LIRGs (except perhaps in ULIRGs). Such quiescent, cool gas at lower densities
can thus contribute significantly to the velocity-integrated flux densities of
the CO $J_{\rm up}\leq3$ lines (as well as to the \ci line of atomic carbon)
increasing the source counts in line surveys that would include those
transitions. Therefore, in addition to the emergent CO line flux computed for
the HCN-bright dense star-forming phase we also include an estimate for the
low-{\it J} CO emission from this cold, quiescent phase.

\subsection{The cold, quiescent gas phase}
\label{phase2}

In the local Universe the star formation `mode' seems to be very well (i.e.
uniquely) indicated by the fraction of the total gas reservoir residing in the
dense phase: $\xi_{\rm SF} = M_{\rm dense}(\rm H_2)/M_{\rm tot}(\rm H_2)$.
Local studies suggest typical values of $\xi_{\rm SF}\sim0.25$ (and even
$\ga $0.50) for merger-driven starburst systems (Solomon et al.\ 1992a; 
Gao \& Solomon 2004; Papadopoulos et al. 2012) and $\xi_{\rm SF}\sim0.025$
 for secular activity in `quiescent' systems, with
the complete star-forming population potentially forming a continuum of
$\xi_{\rm SF}$ (cf. Daddi et al.\ 2010). Therefore, an estimate of the mass of
the quiescent, non-star-forming gas phase can be calculated by:

\begin{equation}
	M_{\rm quiescent}({\rm H_2}) = M_{\rm dense}({\rm H_2})\left(\frac{1-\xi_{\rm SF}}{\xi_{\rm SF}}\right)
\end{equation}

\noindent In the spirit of our `minimal' approach, the most conservative
estimate of $M_{\rm quiescent}(\rm H_2)$ is given by assuming {\it all
galaxies} are in the burst mode, with $\xi_{\rm SF}=0.25$, such that $M_{\rm
quiescent}({\rm H_2}) = 3\times M_{{\rm dense}}({\rm H_2})$. Armed with this
mass estimate, we can assign CO luminosities by assuming a fixed,
representative SLED for this phase.

We assume the `coldest' known CO SLED in the local Universe, which is found in
cold quiescent clouds in M\,31 with CO line ratios: $r_{\rm 21}=0.42\pm 0.10$,
$r_{\rm 32}=0.14\pm 0.04$, $R_{\rm 12/13}{\,J(1\rightarrow0)}={\rm
^{12}CO/^{13}CO}=9\pm 2$ (Allen et al.\ 1995; Loinard et al.\ 1995, 1996;
Loinard \& Allen\ 1998; Israel et al.\ 1998 and references therein). For the
\ci line emission in the same environments $r_{\rm [CI]/CO}{\rm (1-0)}=0.1\pm
0.02$ (Wilson 1997; Israel et al. 1998), while COBE determined the
[CI]($2\rightarrow1$)/($1\rightarrow0$) line ratio (difficult to obtain from
the ground) in the similar environment of the outer Galaxy disk to be $r_{\rm
[CI]}(2\rightarrow1/1\rightarrow0)=0.22\pm 0.09$ (Fixsen et al. 1999). We then
use our LVG radiative transfer model, constrained by the available CO,
$^{13}$CO line ratios and $K_{\rm vir}$$\sim $1 to find the corresponding
average ISM states, higher-{\it J} CO line luminosities, and compute the
$X_{\rm CO}$ factor normalizing the cold CO SLED in terms of H$_2$ molecular
gas mass. We find typical virial LVG solutions with $T_{\rm kin}=(10-15)$\,K,
and $X_{\rm CO}$$\sim $5\,$ M_{\odot}$(K\,km\,s$^{-1}$\,pc$^2$)$^{-1}$ (i.e.
typical of cold GMCs in the Galaxy) and CO line ratios of $r_{\rm
21}=0.45$--$0.55$, $r_{\rm 32}=0.12-0.13$, $r_{\rm 43}=0.09-0.011$ and $r_{\rm
(J+1)-J}$$\la $$10^{-4}$ for $(J+1)$$>$4. We plot the additional cold
component of the quiescent gas phase in Figure\ 1. In Figure\ 2 we present a
schematic plot indicating the observed frequency and atmospheric windows of
the principle emission lines we consider in this work.

\subsection{Far-infrared cooling lines}

The dominant coolant for the warm ($T<1000$\,K) neutral ISM is the
far-infrared (FIR) fine structure line [C{\sc ii}]\,158$\mu$m, carrying up to
1\% of the total FIR output. When redshifted beyond $z>1$ this line will be an
important contributor to detections in the submillimeter bands, and therefore
important to consider in our model.

We adopt an semi-empirical prescription for estimating the luminosity of \cii
based on the compendium of {\it Infrared Space Observatory (ISO)} Long
Wavelength Spectrometer (LWS) observations of 227 local
($cz\lesssim10^4$\,km\,s$^{-1}$) galaxies by Brauher, Dale \& Helou\ (2008).
These authors note the scalings between the principle FIR fine structure line
luminosities and FIR luminosity (here we consider the {\it IRAS} FIR
luminosity to be equivalent to our assumed total infrared luminosity) of
galaxies, as well as the far-infrared color \fircolour (sensitive to the
intensity of the underlying radiation field heating the dust; Stacey et al.\
2010). For example, the ratio $L_{\rm [CII]}/L_{\rm FIR}$ drops from 1\% to
0.1\% with increasing (hotter) \fircolour. This could reflect a harder and
more intense far-UV field that results in conditions less efficient at
producing C$^+$ (Malhotra et al.\ 2001), or an extra contribution to the FIR
continuum from regions not associated with the PDRs (Luhman et al.\ 2003;
Stacey et al.\ 2010). This phenomenon has become known as the [C{\sc ii}]
deficit, although it is not clear whether it is a true deficit.

To evaluate the relevant scalings between line luminosities and IR luminosity,
we only consider galaxies in the Brauher, Dale \& Helou compendium with
nuclear regions classed as `star-forming' that were unresolved by the LWS beam
(D.\ Dale, private communication). Fitting a linear trend to the data, we find
the scaling:

 \begin{equation}\log \left(L_{\rm [CII]}/L_{\rm FIR}\right) = (-0.74\pm0.39)
\left(f_{60}/f_{100}\right) - (2.08\pm0.25).\end{equation}

The uncertainties are the 1$\sigma$ errors in the formal fit and reflect the
large scatter in the data. Although the strength of the \cii line varies by a
factor $\sim$10, and can contribute up to $\sim$1\% of the FIR luminosity, in
the spirit of our minimalist approach, we adopt a conservative scaling that
predicts a modest emergent $L_{\rm [CII]}$ for a galaxy with $L_{\rm
IR}$. We therefore adopt \fircolour$=1$, which yields $\log \left(L_{\rm
[CII]}/L_{\rm FIR} \right)={-2.8}$. While this value is a conservative
estimate for local galaxies, is it appropriate for (on average higher
luminosity) high-redshift star-forming populations? Given the high luminosity
of the line, there is a growing sample of \cii detections in ULIRG-class
systems. For example, Stacey et al.\ (2010) find a an average ratio of $\log
\left(L_{\rm [CII]}/L_{\rm FIR}\right) = -2.5$ for (star formation dominated)
ULIRGs at $1\lesssim z \lesssim 2$; we therefore consider our canonical value
based on the extreme tail of local star-forming galaxies suitable for
application in our model.

\subsubsection{The \cii line as tracer of CO-dark {\rm H$_2$}  at high redshifts}

The presence of a CO-deficient and even CO-dark molecular gas reservoir is
expected in globally metal-poor systems such as local dwarf galaxies as well
as Lyman-break galaxies (LBGs) at high redshifts, the result of efficient CO
dissociation and a strongly self-shielding H$_2$ (e.g.\ Pak et al.\ 1998).
Such a phase is even expected in typical local spiral disks at large
galactocentric distances as a result of metallicity gradients (Papadopoulos et
al.\ 2002). This very much inhibits the detection of such H$_2$ gas reservoirs
via the workhorse CO lines but at the same time anticipates very bright \cii
emission. In local dwarf galaxies the latter revealed $\sim $10--100 times
larger molecular gas mass than that inferred by CO (Madden et al.\ 1997).

Similarly strong \cii emission is expected for high-{\it z} systems like LBGs
which are very difficult to detect in CO lines (see Baker et al.\ 2004, Coppin
et al.\ 2007 for the only such examples), while early evidence seems to
corroborate this for other types of high-{\it z} systems (Maiolino et al.\
2009). A general \cii emission enhancement at high redshifts with respect to
that expected from local template systems with similar infrared luminosities
can be a result of a general evolutionary trend towards more metal-poor gas
reservoirs at earlier epochs, and could much enhance the potential of the \cii
as a galaxy redshift survey tool. We caution though that its emission, once
detected, will be much harder than, for example, \ci to interpret solely 
in terms of total molecular gas mass as the \cii line will contain
 significant contributions from ionized and neutral hydrogen 
(e.g.\ Madden et al.\ 1997). These non-H$_2$ contributions to \cii
line emission will be hard to correct for at high redshifts, where
they may actually  be boosted in metal-poor environments and/or
strong average far-UV radiation fields of nascent starbursts.

  To explore these possibilities we assume a population of objects with
  $L_{\rm [CII]}/L_{\rm IR}=10^{-2}$, a value typical for the LMC and IC\,10
  (Madden et al.\ 1997). At such levels the \cii line of LIRG-class systems
  can be detected right out to the epoch of re-ionisation ($z\sim10$) with
  ALMA in bands 4--9, within 8\,hrs of full aperture synthesis observations
  (assuming one bin channel of 300\,km\,s$^{-1}$, see \S5.1). The
  evolutionary track of the SFR of such a putative galaxy population with a
  nearly CO-dark ISM is unknown, leaving the line number counts uncertain,
  however we discuss the potential for blind discovery of such systems
  assuming their abundance and evolution is identical to that of LBGs in
  \S5.1.1.

\subsection{Calculating line luminosities}

For the various computations and transformations between line luminosity
types, and for linking the latter to observed velocity-integrated line fluxes
we use standard relations,

	\begin{align}
\nonumber 	L'_x & =\int  _{\Delta  V}\int  _{A_s} T_{b,x}\,da\,dV \\
	 & = \frac{c^2}{2k_B\nu^2 _{\rm x,rest}}\left(\frac{D^2 _L}{1+z}\right) \int _{\Delta V} S_{\nu}\,dV,
	\end{align}

\noindent where $T_{\rm b,x}$ is the rest-frame brightness temperature of the
line, $\Delta V$, $A_{\rm s}$ are the line FWZI and source area respectively.
After substituting astrophysical units this yields,

	\begin{align}
\nonumber 	 L'_x=	 3.25\times 10^7 & \left[\frac{D^2 _{L}({\rm Mpc})}{1+z}\right]\left(\frac{\nu_{\rm x,rest}}{\rm GHz}\right)^{-2} \\
 &	\times \left[\frac{\int _{\Delta V} S_{\nu }\,dV}{\rm Jy\,km\,s^{-1}}\right]\,{\rm K\,km\,s^{-1}\,pc^2},
	\end{align}

\noindent where $D_{L}$ is the luminosity distance, and $\nu _{\rm x,rest}$ is
the rest frame line frequency. The conversion to traditional luminosity units
($L_{\odot}$), used for the total line luminosities ($L_{x}$=$\int
L_{\nu}\,d\nu$) in CO SLEDs, can be made using

	\begin{align}
\nonumber 	L_x & =\frac{8\pi k_B\nu ^3 _{\rm x,rest}}{c^3}\,L' _x\\
	& =3.18\times 10^4\left(\frac{\nu _{\rm x,rest}}{\rm 100\,GHz}\right)^3
	\left[\frac{L' _x}{10^{9}\,{\rm K\,km\,s^{-1}\,pc}^2}\right]L_{\odot}.
	\label{eqn:sol}
	\end{align}

Note that integrated line fluxes are related to the commonly used
$S\Delta V$ integrated flux density units by:

\begin{equation}\left(\frac{3\times10^{-4}}{\nu_{\rm obs}\,\,{\rm GHz}}\right)\times\left(
\frac{F}{{\rm 10^{-26}\,W\,m^{-2}}} \right) = \left(\frac{S\Delta V}{\rm
Jy\,km\,s^{-1}}\right).\end{equation}

	\begin{table}
		\caption{Normalized emergent CO and HCN SLEDs}\vspace{-0.5cm}
	\begin{center}
		\begin{tabular}{r@{$\rightarrow$}lccc}
			\hline
			\multicolumn{2}{r}{Transition} &
			\multicolumn{3}{c}{$r_{\rm J+1,J}$$^{\rm a}$} \cr
		   \multicolumn{2}{r}{\phantom{Transition}} & CO
		   (dense) & CO (quiescent) & HCN\cr
			\hline
		1&0 & 1.00 &1.00 & 1.00 \cr
		2&1 & 0.95 &0.50 & 0.75\cr
		3&2 & 0.88 & 0.13 & 0.40\cr
		4&3 & 0.82 & 0.09 & 0.15 \cr
		5&4 & 0.70--0.75 & 10$^{-4}$ & 0.05\cr
		6&5 & 0.64--0.70 & \cr		
		7&6 & 0.53--0.60\cr
		8&7 & 0.31--0.42\cr
		9&8 & 0.10--0.20\cr
		10&9& 0.01--0.04\cr 
		\hline
		\multicolumn{4}{l}{$^{\rm a}$$r_{\rm J+1,J} = L'_{{\rm J+1\rightarrow J}}/L'_{{\rm 1\rightarrow0}}$}
			\end{tabular}
	\end{center}
	\vspace*{-0.28cm}
	\label{tab:sled}
	\end{table}

\subsection{The normalization of dense molecular gas SLEDs}

The final remaining step for calculating the emergent molecular line emission
from a star-forming galaxy is to normalize the SLEDs according to the infrared
luminosity (i.e.\ SFR) of the system. Recent studies of star formation
feedback suggest a maximum $\epsilon_{\star}= L_{\rm IR}/M_{\rm dense}({\rm
H}_2)\sim 500$\,$ (L_{\odot}/M_{\odot})$ for the dense and warm gas $M_{\rm
dense}({\rm H}_2)$ near star forming sites in galaxies as a result of strong
radiation pressure from the nascent O, B star clusters onto the concomitant
dust of the accreted gas fuelling these sites (Scoville\ 2004; Thompson et
al.\ 2005; Thompson\ 2009). Thus, provided that average dust properties (e.g.\
its effective radiative absorption coefficient per unit mass) remain similar
in metal-rich star-forming systems such as LIRGs, a near-constant
$\epsilon_{\star}$ is expected for the dense star-forming gas. A value of
$\epsilon_{\star}\sim 500$\,$(L_{\odot}/M_{\odot})$ is actually measured in
individual star-forming sites of spiral disks such as M\,51 and entire
starbursts such as Arp\,220 (Scoville\ 2004), while $\sim$($440\pm 100$)\,$
(L_{\odot}/M_{\odot})$ is obtained for CS-bright star-forming cores in the
Galaxy (Shirley et al.\ 2003). It must be noted however that the intermittency
expected for galaxy-sized molecular gas reservoirs (i.e.\ at any given epoch
of a galaxy's evolution some dense gas regions will be forming stars while
others will not) can lower the global $\epsilon_{\star}$ to
$\sim$\nicefrac{1}{3}--\nicefrac{1}{2} of the Eddington value (Andrews \&
Thompson~2011).

Similar $\epsilon_{\star}$ values can be obtained without explicit use of the
Eddington limit (and the detailed dust properties it entails), but from the
typical $L_{\star}/M_{\rm new,\star}$ in young starbursts where $M_{\rm
new,\star}$ is the mass of the new stars and $L_{\star}$ their bolometric
luminosity ($\sim$$L_{\rm IR}$ for the deeply dust-enshrouded star-forming sites). For
$\epsilon _{\rm SF,c}$=$ M_{\rm new,\star}/[M_{\rm new,\star}+M_{\star}({\rm
H}_2)]$ as the star formation efficiency (SFE) of the dense gas regions where
the new stars form is:

	\begin{equation}
	\epsilon_{\star}=\frac{\epsilon_{\rm SF,c}}{1-\epsilon
	_{\rm SF,c}}\left(\frac{L_{\rm IR}}{M_{\rm new,\star}}\right).
	\label{eqn:sfe}
	\end{equation}

\noindent For $\epsilon_{\rm SF,c}$$\sim $0.3--0.5 typical for dense star-forming
regions, and $L_{\rm IR}/M_{\rm new,\star}=$300--400\,$(M_{\odot}/L_{\odot})$
(Downes \& Solomon 1998 and references therein), equation\ 8 yields
$\epsilon_{\star}$$\sim $130--400\,$(M_{\odot}/L_{\odot})$. Here we choose
$\epsilon_{\star}=$250\,$(L_{\odot}/M_{\odot})$, close to the average values
yielded by equation 8, and the black body limit deduced for the
compact CO line emission concomitant with an optically thick ($\tau_{\rm
100\,\mu m}$$>$1) dust emission seen in ULIRGs (Solomon et al.\ 1997).

Thus Eddington-limited (i.e.\ radiation pressure limited) star formation in
LIRGs sets a near-constant mass normalization of the dense star-forming gas
phase using the star formation driven infrared luminosity and thus allows a
$M_{\rm dense}({\rm H}_2)$/SFR value to be deduced. This can then be used to
set the absolute scale of the emergent minimal CO SLEDs of the star-forming
gas phase using the SFR($z$) of an input galaxy evolution model, since it is

 \begin{align}\nonumber  L'_{\rm CO}(1-0) & =M_{\rm dense}({\rm H_2})/X_{\rm CO}\\
\nonumber  & 	= L_{\rm IR}/\epsilon_{*}X_{\rm CO} \\
 &	= k_{\rm sfr} {\rm SFR}/\epsilon_{*}X_{\rm CO}
\end{align}

which can be converted to solar units using equation\ 6, and with an identical
relation applying to the \hcni line luminosity but utilizing the corresponding
$ X_{\rm HCN}$ values. 

In Figure\ 3 we present the observed CO line fluxes predicted for both the
virial and super-virial models for a range of $L_{\rm IR}$ and redshift; this
can be used as a `ready reckoner' to estimate the CO line flux for a given
galaxy where some estimate of the infrared luminosity is known. As noted in
\S3.1, the LVG solutions for the dense gas phase are $ X_{\rm CO}$=\{3,
9\}\,$M_{\odot}$(K\,km\,s$^{-1}$\,pc$^2$)$^{-1}$, and $ X_{\rm HCN}$=\{9,
27\}\,$M_{\odot}$(K\,km\,s$^{-1}$\,pc$^2$)$^{-1}$ for the super-virial and
virial cases. Given the well-excited CO SLEDs expected for the dense
star-forming gas phase (Figure\ 1), their normalization to the dense gas mass,
namely the $ X_{\rm CO}$ factor, determines to a great degree the `visibility'
of these SLEDs in the distant Universe. For a given amount of dense gas lower
$X_{\rm CO}$, $X_{\rm HCN}$ values than the adopted ones correspond to
brighter CO and HCN SLEDs though we consider them unlikely for the
self-gravitating or only modestly unbound dense star-forming gas in LIRGs.

The LVG-derived CO line ratios (Table\ 1) can then be used to derive the
expected fluxes for the other CO transitions from the SFR-normalized $L'_{\rm
CO(1-0)}$. For systems with lower metallicities we assume proportionally less
dust per molecular gas, and thus the Eddington limit becomes
$\epsilon_{\star}=250(Z/Z_\odot)^{-1}$\,$ (L_{\odot}/M_{\odot})$, where $Z$ is
the metallicity. Finally we note that while the dense HCN-bright gas fuelling
star formation will be often a small fraction of the total molecular gas mass
in galaxies it is nevertheless the only phase for which the frequently-used
gas consumption timescale $\tau_{\rm cons}$=$M_{\rm dense}({\rm H}_2)$/SFR may
have its intended physical meaning as the duration of an observed
star formation event. Strong star-formation and/or AGN feedback will almost
certainly modify the `consumption' timescale especially for the extended, less
dense, quiescent molecular gas by inducing powerful outflows (Sakamoto et al.\
2009; Feruglio et al.\ 2010; Chung et al.\ 2011), or by fully dissociating
large fractions of H$_2$ gas mass towards warmer and diffuse phases unsuitable
for star formation such as Cold Neutral Medium (CNM) and Warm Neutral Medium
(WNM) H{\sc i} gas (Pelupessy et al.\ 2006).

Lower metallicities will not have any effect on the emergent CO SLED shapes of
the dense star forming gas phase in galaxies as long as the optical depths of
the CO lines remain significant. Nevertheless the CO-bright part of individual
H$_2$ clouds will shrink, leaving behind \cii and \ci bright H$_2$ gas
(Bolatto et al.\ 1999 [Figure\ 1]; Pak et al.\ 1998). Thus the expected first
order effect of lower metallicities would be to lower all CO line luminosities
per H$_2$ mass, while boosting up the \cii and \ci line luminosities. The
latter may be crucial in the detection of metal-poor galaxies at high
redshifts (see section 3.3.1).

\begin{figure}[t]
\centerline{\includegraphics[width=0.5\textwidth,angle=-90]{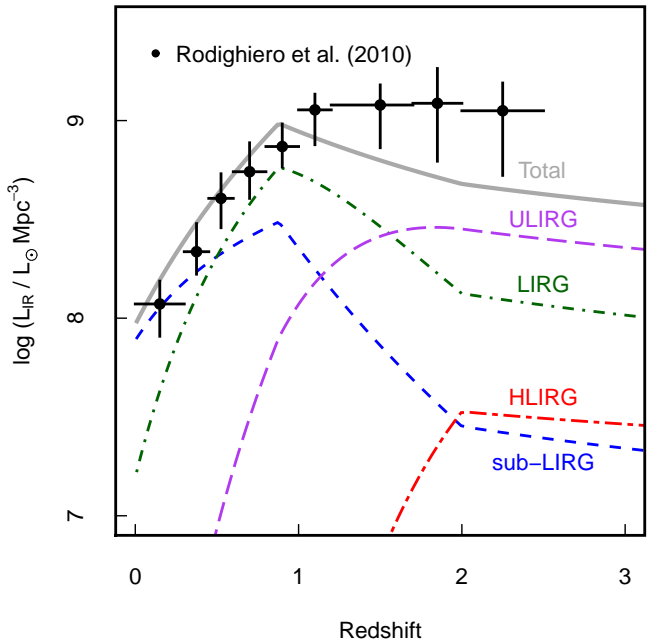}}
\caption{Model of the evolution of the infrared luminosity density,
figure adapted from Figure\ 12 of B\'ethermin et al.\ (2011). Breaking
down the total contribution into sub-LIRG ($L_{\rm
IR}<10^{11}L_\odot$), LIRG ($L_{\rm IR}=10^{11-12}L_\odot$), ULIRG
($L_{\rm IR}=10^{12-13}L_\odot$) and HLIRG ($L_{\rm
IR}>10^{13}L_\odot$) clearly shows how the dominance changes from
low-luminosity spiral galaxies in the local Universe, through LIRGs at
the peak of the luminosity density and to ULIRGs at $z>2$. We use the
model of B\'ethermin et al. as a basis for predicting the number
density of star-forming galaxies across cosmic time. The points show
the measurements of Rodighiero et al.\ (2010) which are based on {\it
Herschel} PACS observations of the GOODS-N field. This figure also
illustrates the basic form of the model: rapid evolution in both space
density and characteristic luminosity to $z\sim0.9$, followed by two
phases of negative evolution to high-{\it z}.} \label{fig:lirdensity}
\end{figure}

\begin{figure*}
\centerline{\includegraphics[height=\textwidth,angle=-90]{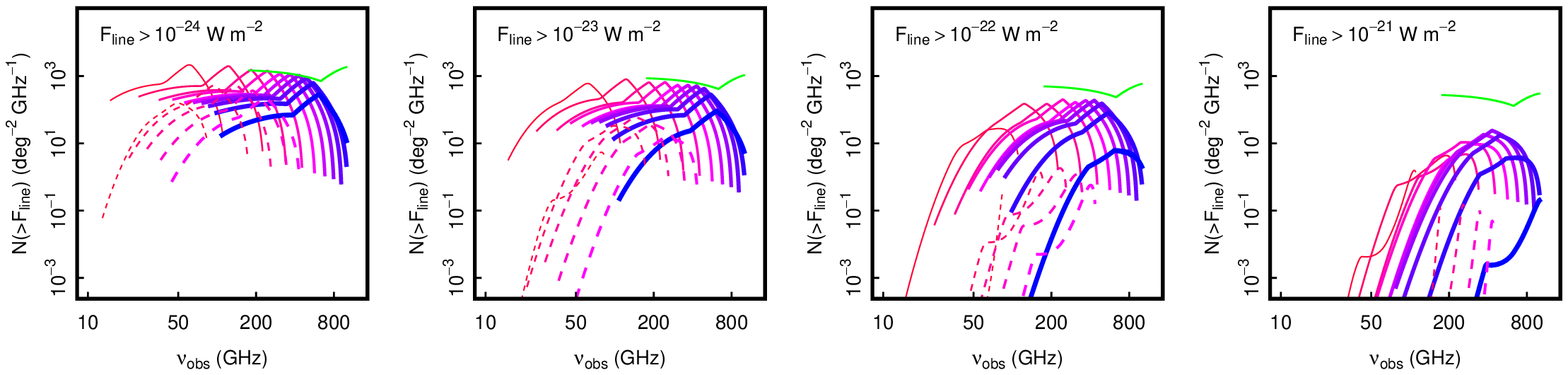}}
\centerline{\includegraphics[height=\textwidth,angle=-90]{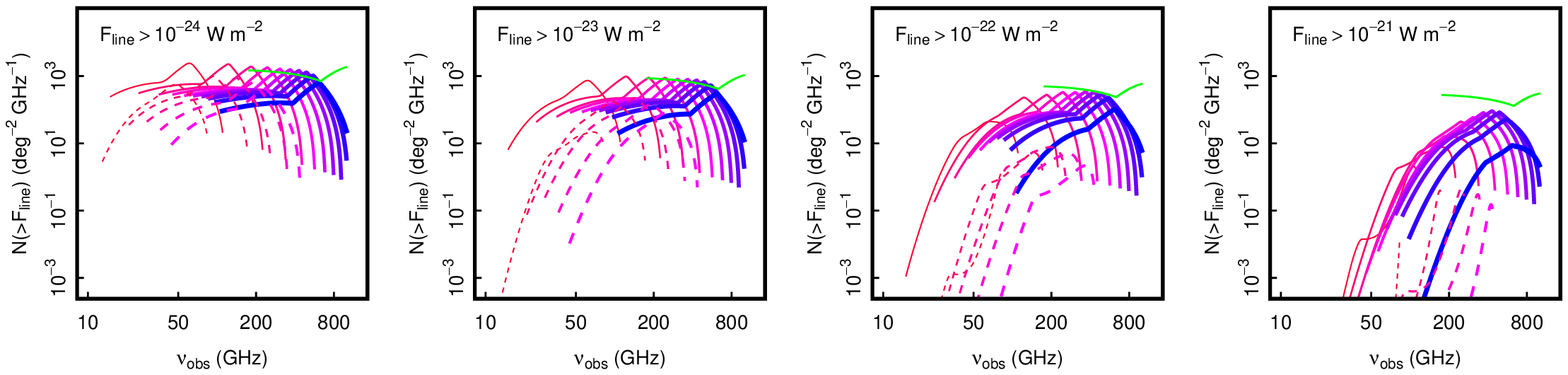}}
\caption{Counts of galaxies with CO\,$J_{\rm up}\leq 10$ (solid lines,
blue/thick to red/thin for increasing {\it J}), HCN\,$J_{\rm up}\leq
5$ (dashed lines blue/thick to red/thin for increasing {\it J}), \cii (green line) and $F_{\rm line}$$>$$10^{-24}$--$10^{-21}$\,W\,m$^{-2}$ as a function of
observing frequency for the virial {\it (top row)} and super-virial SLED
models {\it (bottom row)}. We calculated the integrated counts for each line
in a sliding 8\,GHz wide window, and therefore quote the counts as the average
per unit bandpass. The sharp cut-offs at the highest frequencies of each line
correspond to $z=0$, and the predictions terminate at frequencies
corresponding to $z=10$ in this figure. The distinctive shape of the counts when presented in
this way is simply a reflection of the form of the evolution of the IR
luminosity function, which has a peak of activity at $z\sim1$--$2$, followed
by weak negative evolution to high-$z$ (see \S4). }\label{fig:ghz}
\end{figure*}

\subsection{\ci and HCN lines: a new promising avenue towards total gas mass
and the star formation mode}

The submillimeter lines of atomic carbon, and especially \ci at 492\,GHz have
been proposed as powerful alternatives to the \coi or \coii lines as total
molecular gas mass tracers (Papadopoulos et al.\ 2004), and have been shown to
work to that effect in local LIRGs (Papadopoulos \& Greve\ 2004). The \ci line
may actually be a {\it better} tracer of total molecular gas mass than \coi
(Papadopoulos et al.\ 2004), taking advantage of the positive {\it
k}-correction with respect to the \coi line (while the \coiv line at similar
frequency to \ci is a poor total molecular gas mass tracer as it is tied to
the star-forming gas phase). Moreover \ci is optically thin (thus it does not
need `$X$' factors to trace mass), has a simple partition function, is easily
excitable for the bulk of the H$_2$ gas mass in galaxies, and of course
remains accessible to ALMA for a much wider range of redshifts than \coi or
\coii (Figure\ 2). When combined with a tracer of the dense, star-forming gas
phase, combinations of, e.g.\ \hcni and \ci could be a powerful tracer of the
star formation `mode' of a galaxy, as traced by $M_{\rm dense}({\rm
H}_2)/M_{\rm total}({\rm H}_2)$. We discuss the prospect of practical
surveys of the star formation mode in an accompanying Paper II (Papadopoulos
\& Geach\ 2012).

\section{Galaxy number counts model}

In order to use the line emission model to predict the number counts in a
blind molecular line survey, we require a framework for the evolution of the
abundance of galaxies as a function of infrared luminosities (i.e.\ the SFR
history traced by the evolution of the infrared luminosity density). This can
then be used to derive the corresponding evolution of $M_{\rm dense}({\rm
H_2})$, and thus the emergent molecular line emission.

We use the `backwards evolution' parametric model of B\'ethermin et al.\
(2011), which is a phenomenological model of the evolution of the bolometric
luminosity function, based on the philosophy of Lagache et al. (2003), but
updated to fit the latest empirical results from large area infrared and
sub-mm surveys, including recent results from {\it Herschel}. B\'ethermin et
al.\ (2011) employ a three-step evolution of the IR luminosity function, which
evolves as $(1+z)^Q$ in $\phi_\star$ and $L_\star$, with two redshift breaks
(at $z=0.89$ and $z=2$) where the exponents of the evolution coefficients
change. In summary, the model can be described as a steep rise in the IR
luminosity density out to the first break, followed by a rapid negative
evolution to $z=2$ (i.e.\ there is a `burst' of activity at $z\sim1$--$2$),
followed by weak negative evolution to high-$z$. The model is only empirically
constrained to $z\sim3$, and so we caution the reader that extrapolations to
higher redshifts, as we present here, are inherently uncertain. We examine the
impact of an order-of-magnitude change in the infrared luminosity density
evolution at $z>3$ on our predicted counts in \S5.1.1, and we acknowledge that
any predictions we make for the abundance of gas reservoirs at very high
redshifts are rather speculative; this was, in part, our motivation for a
`minimal' ISM model for the number counts.

The infrared luminosity density evolution model is successful at re-producing
the differential counts across the full range of typical bandpasses,
24--850$\mu$m, as well as the observed evolution of the infrared luminosity
function. The evolution of the infrared luminosity density broken into
contributions from sub-LIRG ($L_{\rm IR}<10^{11}L_\odot$), LIRG ($L_{\rm
IR}=10^{11-12}L_\odot$), ULIRG ($L_{\rm IR}=10^{12-13}L_\odot$) and HLIRG
($L_{\rm IR}>10^{13}L_\odot$) classes is shown in Figure\,4, compared to the
latest measurements of the total infrared luminosity density from {\it
Herschel} (Rodighiero et al.\ 2010). Beyond $z\sim1$ the model slightly
under-predicts the luminosity density (although the observational constraints
become more uncertain at this epoch), indicating that the model could be
considered a conservative estimate of the galaxy counts at high-$z$.

\begin{table*}
	\caption{Predicted number counts of molecular emission lines}
	\vspace{-0.5cm}
\begin{center}
%\tiny
\begin{tabular}{lcr@{.}lcccccr@{.}lr@{.}lc}
\hline
\hline
	 	Band & $\nu_{\rm obs}$ & \multicolumn{2}{c}{FoV$^{\rm a}$} & rms$^{\rm c}$ & $\log F_{\rm 1\,hour}$ & $\log F_{\rm opt}$ & $N(>$$F_{\rm 1\,hour})$ & $N(>$$F_{\rm opt})$ &  \multicolumn{2}{c}{1hr rate} & \multicolumn{2}{c}{Optimal rate} & Serendipitous\cr
	 & GHz & \multicolumn{2}{c}{arcmin$^2$} & $\mu$Jy\,$\surd$hr &W\,m$^{-2}$ & W\,m$^{-2}$ & deg$^{-2}$ & deg$^{-2}$ & \multicolumn{2}{c}{hour$^{-1}$} & \multicolumn{2}{c}{hour$^{-1}$} &  10\,hr$^{\rm e}$\cr
		\hline

\cr

\multicolumn{14}{c}{Virial SLED model}\cr
\cr
\hline
\cr

MeerKAT$^{\rm b}$ & 10 & 55&44 & 38 & $-$22.7 & $-$23.4 & $<$1 & 92 & 0&011  & 0&046 & 0.39\cr
SKA K$^{\rm b}$ & 18 & 11&16 & 1 & $-$23.8 & $-$22.8 & 2215 & 122 & 6&867  & 35&282 & 14.93\cr
ALMA Band 3 & 103 & 0&83 & 77 & $-$21.4 & $-$21.5 & 174 & 231 & 0&040  & 0&041 & 0.28\cr
ALMA Band 4 & 147 & 0&40 & 77 & $-$21.2 & $-$20.0 & 629 & 33 & 0&070  & 1&099 & 0.24\cr
ALMA Band 5 & 163 & 0&24 & 86 & $-$21.2 & $-$20.0 & 761 & 69 & 0&050  & 0&881 & 0.15\cr
ALMA Band 6 & 212 & 0&14 & 75 & $-$21.1 & $-$21.1 & 946 & 1033 & 0&037  & 0&034 & 0.10\cr
ALMA Band 7 & 278 & 0&08 & 87 & $-$20.9 & $-$21.0 & 911 & 1137 & 0&020  & 0&015 & 0.05\cr
ALMA Band 8 & 406 & 0&04 & 288 & $-$20.2 & $-$21.0 & 383 & 1216 & \multicolumn{2}{c}{\ldots}  & \multicolumn{2}{c}{\ldots}& \ldots\cr
ALMA Band 9 & 668 & 0&02 & 721 & $-$19.6 & $-$21.0 & 265 & 1026 &\multicolumn{2}{c}{\ldots}  & \multicolumn{2}{c}{\ldots} & \ldots\cr
ALMA Band 10 & 854 & 0&01 & 1526 & $-$19.2 & $-$20.0 & 357 & 922 & \multicolumn{2}{c}{\ldots}  & \multicolumn{2}{c}{\ldots} & \ldots\cr

\hline
\cr
\multicolumn{14}{c}{Super-virial SLED model}\cr
\cr
\hline

MeerKAT$^{\rm b}$ & 10 & 55&44 & 38 & $-$22.7 & $-$23.3 & 1 & 91 & 0&029  & 0&079 & 0.75\cr
SKA K$^{\rm b}$ & 18 & 11&16 & 1 & $-$23.8 & $-$22.7 & 3164 & 135 & 9&811  & 70&894 & 21.34\cr
ALMA Band 3 & 103 & 0&83 & 77 & $-$21.4 & $-$20.1 & 623 & $<$1 & 0&144  & 0&040 & 0.66\cr
ALMA Band 4 & 147 & 0&40 & 77 & $-$21.2 & $-$20.0 & 1341 & 34 & 0&150  & 1&122 & 0.46\cr
ALMA Band 5 & 163 & 0&24 & 86 & $-$21.2 & $-$20.0 & 1730 & 74 & 0&115  & 0&946 & 0.31\cr
ALMA Band 6 & 212 & 0&14 & 75 & $-$21.1 & $-$20.5 & 2195 & 483 & 0&086  & 0&336 & 0.20\cr
ALMA Band 7 & 278 & 0&08 & 87 & $-$20.9 & $-$20.3 & 2176 & 576 & 0&048  & 0&181 & 0.11\cr
ALMA Band 8 & 406 & 0&04 & 288 & $-$20.2 & $-$20.3 & 615 & 703 & 0&007  & 0&006 & 0.02\cr
ALMA Band 9 & 668 & 0&02 & 721 & $-$19.6 & $-$20.2 & 283 & 566 & \multicolumn{2}{c}{\ldots}  & \multicolumn{2}{c}{\ldots} & \ldots\cr
ALMA Band 10 & 854 & 0&01 & 1526 & $-$19.2 & $-$20.0 & 359 & 940 & \multicolumn{2}{c}{\ldots}  & \multicolumn{2}{c}{\ldots} & \ldots\cr

	\hline 
	 \multicolumn{14}{l}{$^{\rm a}$assuming field-of-view is
	 equivalent to full width at half maximum of primary beam}\cr 

\multicolumn{14}{l}{$^{\rm b}$counts per 4\,GHz window}\cr

\multicolumn{14}{l}{$^{\rm c}$1$\sigma$ noise in 300\,km\,s$^{-1}$ channel}\cr

 \multicolumn{14}{l}{$^{\rm d}$time required to detect integrated 300\,km\,s$^{-1}$ line at 5$\sigma$}\cr 

\multicolumn{14}{l}{$^{\rm e}$number of sources detected at $>$5$\sigma$ in
300\,km\,s$^{-1}$ channels in a 10\,hour integration, per field-of-view in a 4
or 8\,GHz bandwidth}\cr 
\multicolumn{14}{l}{`\ldots'\ indicate the
counts are effectively zero for practical purposes
($<$$10^{-3}$\,hr$^{-1}$)}\cr

\end{tabular} 

\end{center}

\label{tab:optimal} \end{table*}

 \begin{figure*}
\centerline{\includegraphics[width=0.95\textwidth,angle=-90]{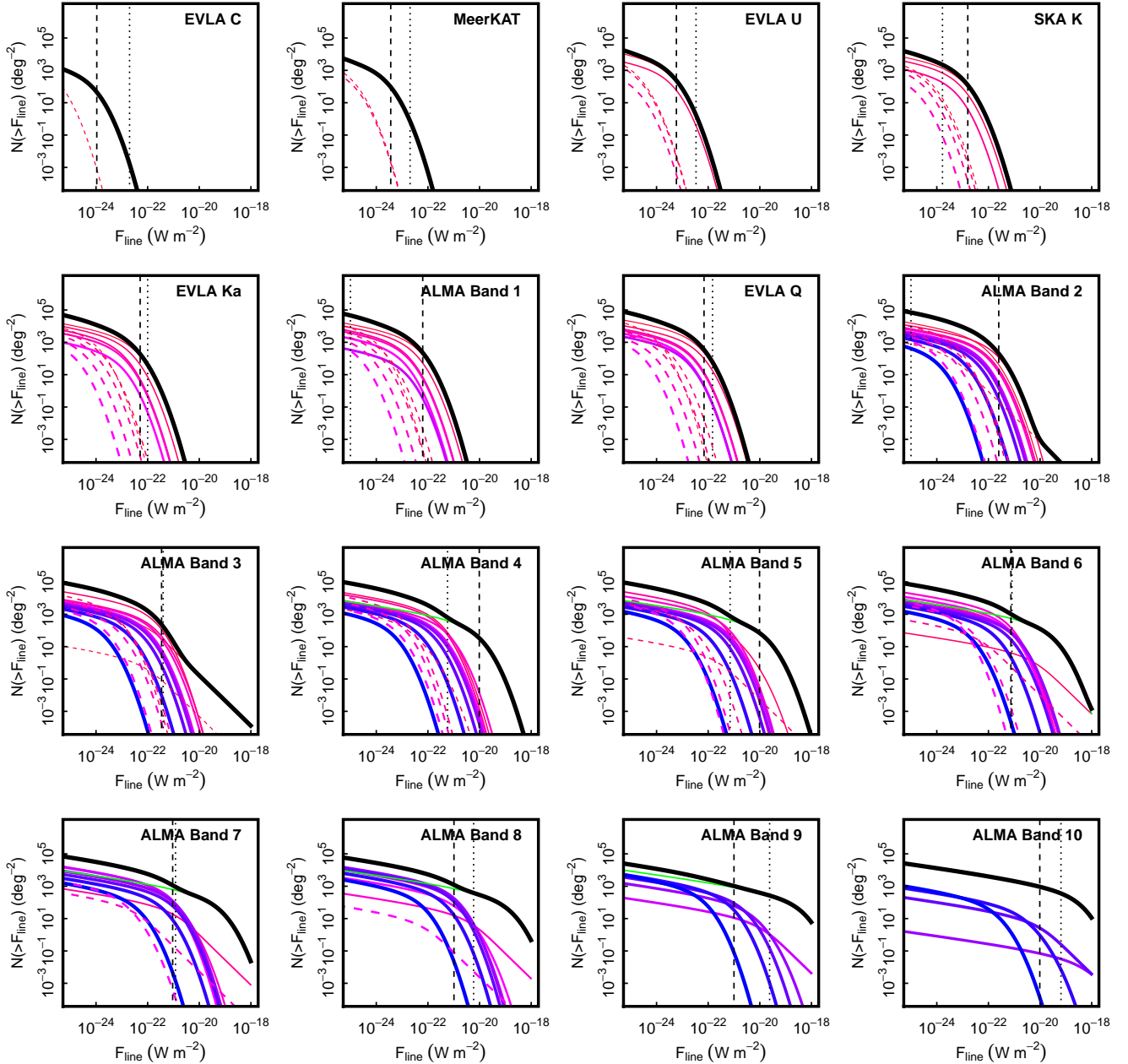}}
\caption{Integrated number counts in our virial SLED model of CO\,$J_{\rm
up}\leq10$, HCN\,$J_{\rm up}\leq5$ and \cii lines in 8\,GHz wide bands at the
central frequencies of the submm--cm bands (although Bands 1 and 2 are
currently hypothetical, pending development). The dashed vertical line
indicates where the slope of the counts rolls through $N\propto F^{-2}$, thus
indicating the flux limit corresponding to an `optimal' survey at a given
frequency (Blain et al.\ 2000) and the vertical dotted line indicates the
5$\sigma$ depth for a 1\,hour integration (this is unknown for ALMA Band 1 and
2).} \label{fig:intcounts} \end{figure*}

 \begin{figure*}
\centerline{\includegraphics[width=0.95\textwidth,angle=-90]{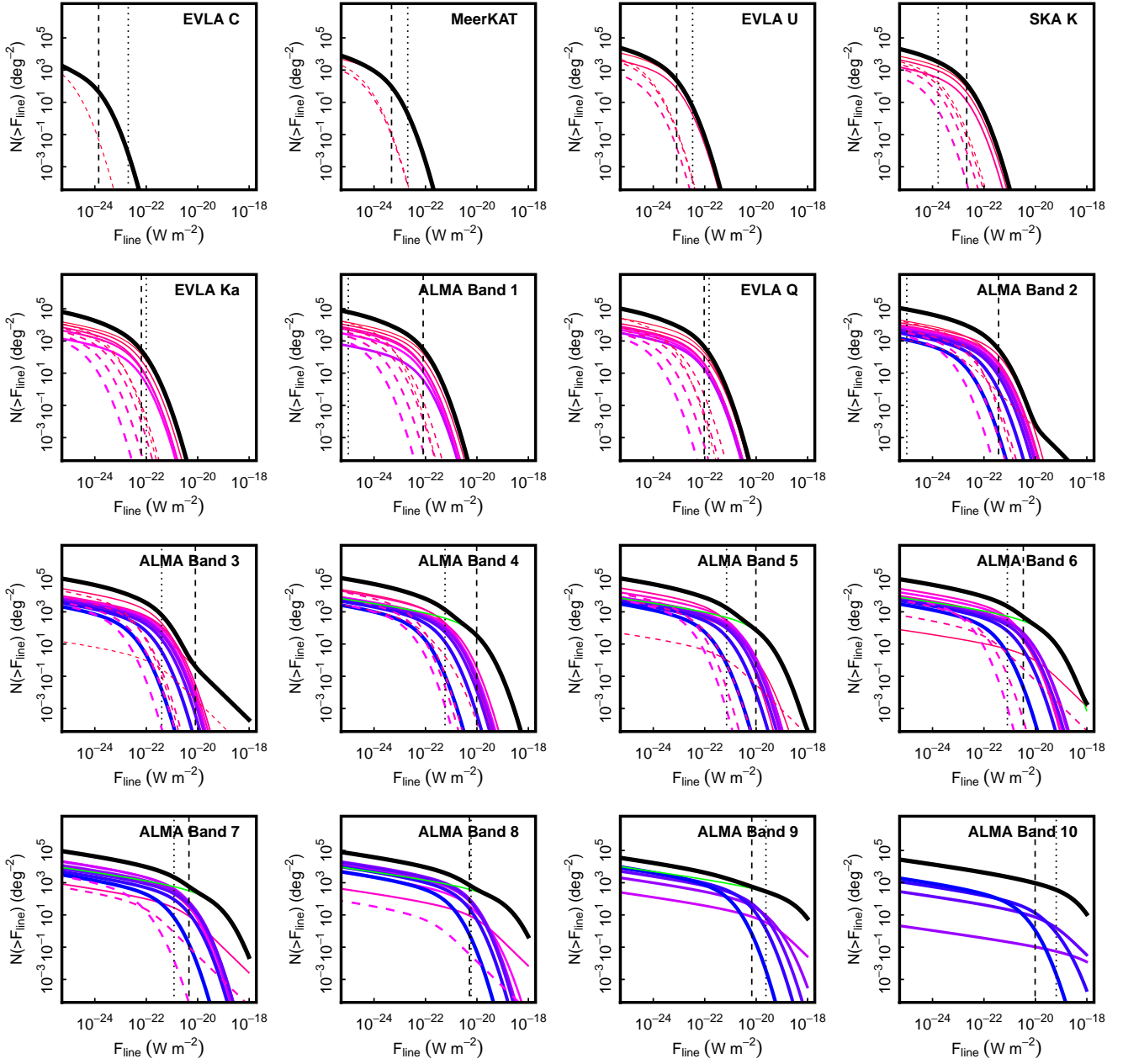}}
\caption{Integrated number counts in our super-virial SLED model,
caption as Figure 6.} \label{fig:intcounts} \end{figure*}

\begin{figure*}
\centerline{\includegraphics[width=0.95\textwidth,angle=-90]{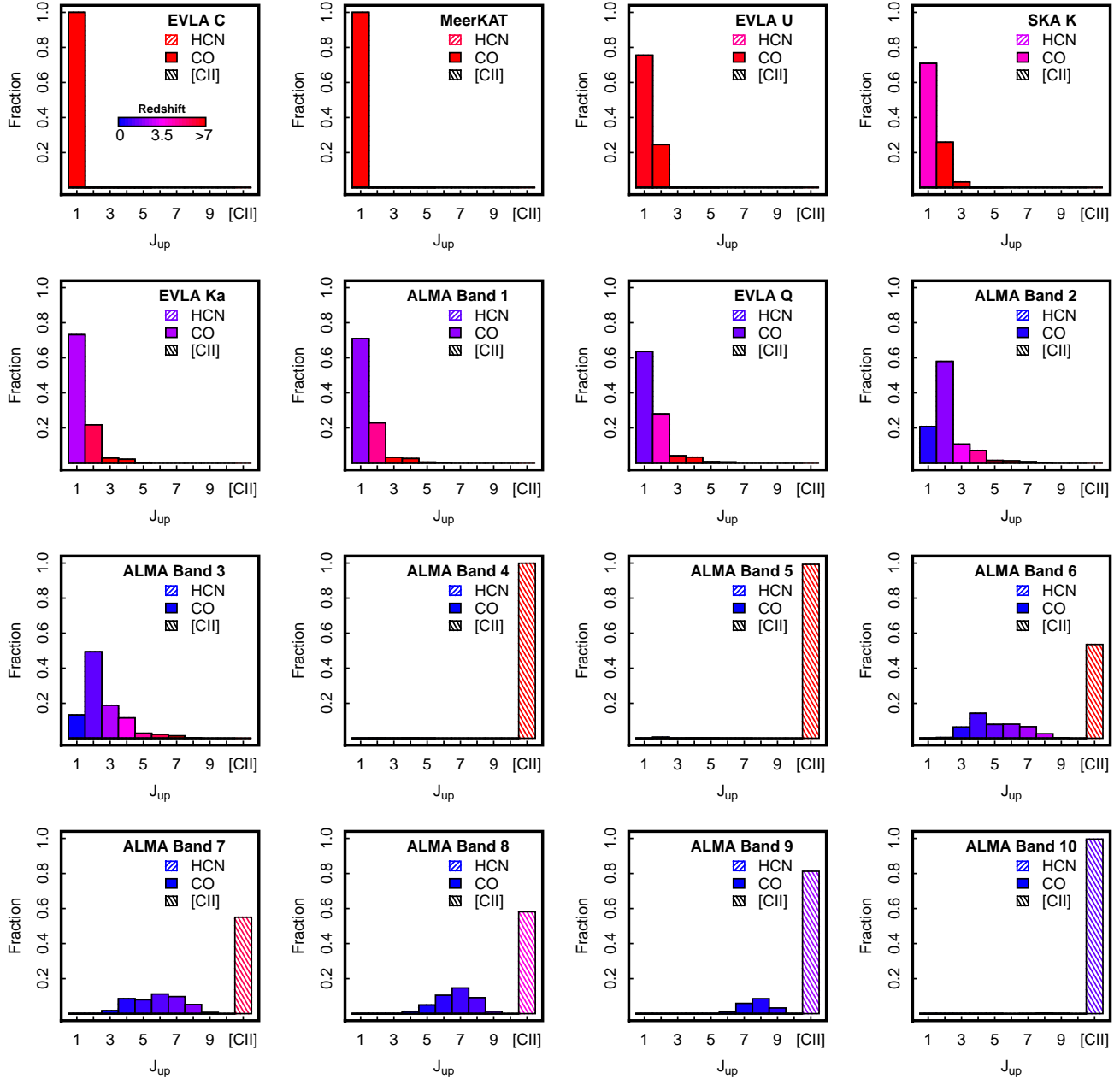}}
\caption{Break-down of line species detected at the optimal flux limit
of each band in the virial model (Fig.\ 6). Bars are colored by the
average redshift of each line, clearly indicating the transition from
the dominance of low-{\it J}/high-{\it z} detections at low
frequencies (accessible with MeerKAT, JVLA and MeerKAT, etc.) through a more
diverse yield of mid-{\it J}/medium-{\it z} lines and \cii dominance
in the ALMA bands. HCN emitters always represent a tiny fraction of
detections in optimal surveys.} \label{fig:breakdown_virial} \end{figure*}

\begin{figure*}
\centerline{\includegraphics[width=0.95\textwidth,angle=-90]{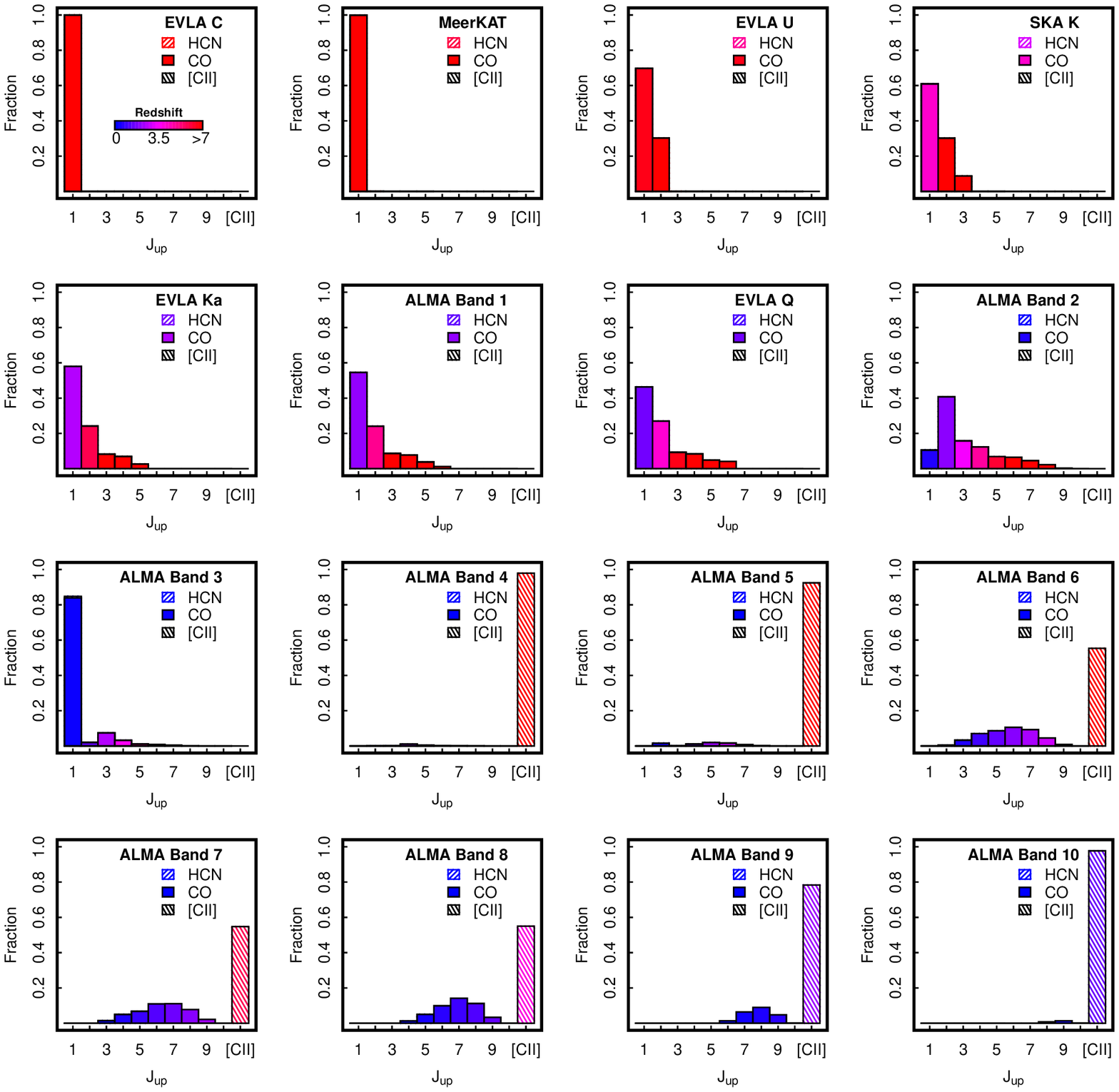}}
\caption{Break-down of line species detected at the optimal flux limit of each band in the super-virial model
(Fig.\ 7). Caption as Figure\ 8.} \label{fig:breakdown_super} \end{figure*}

\section{Results}

\subsection{Minimal integral number counts in a line flux limited survey}

The model of infrared number counts provides the framework on which to predict
the number counts of molecular lines emitted by star-forming galaxies, since
$L_{\rm IR}\propto M_{\rm dense}({\rm H_2})$. We apply the minimal emergent
model of CO emission presented in \S3 to predict $L_{\rm IR}$ for a given line
of integrated flux $F_{\rm line}$ seen at observed frequency $\nu_{\rm obs}$.
The $L_{\rm IR}(z)$ model can then be integrated to predict the lower limit to
the number counts of objects detected in a line flux limited survey across the
full mm--cm regime.

In Figure\ 5 we plot the minimal number counts across $20<\nu_{\rm
obs}<1200$\,GHz for \cii, HCN and CO\,$J_{\rm up}\leq 10$ line emitting
galaxies with $F_{\rm line}>10^{-24}$--$10^{-21}$\,W\,m$^{-2}$, averaged over
an 8\,GHz sliding window. Care will have to be taken to correctly identify
single lines; the most robust strategy would be to detect two or more
molecular lines for an accurate redshift identification. The shape of the
counts of the individual lines reflects the evolution of the underlying
luminosity density described in \S4.

This burst of activity at $z\sim1$ results in a corresponding peak in the line
counts, which is probably artificially sharp in the present model; in reality
we might expect a smoother turn-over (this can be mimicked by averaging the
counts over a wider bandwidth). Nevertheless, this demonstrates the power a
molecular line survey could have as a probe of the evolution of the galaxy
population, as the yield of line emitters in a spectral survey over
sufficiently wide frequency range ($>$100\,GHz) will be a clean and sensitive
tracer of the history of galaxy evolution that is complementary to previous
studies that have focused on the evolution of the SFR and stellar mass
functions.

It will not be possible to fully sample the frequency range shown in Figure\
2. Line searches from the ground are strictly restricted to the frequency
windows dictated by the atmospheric transmission (the most significant for
evolutionary surveys being the telluric feature at $\sim$60\,GHz), and the
availability of receivers capable of detecting the radiation transmitted
through such windows. In the following sections we examine the number counts
expected in practical blind surveys conducted by the latest arrays with the
sensitivity capable of performing them: ALMA, SKA and its pathfinder MeerKAT.
JVLA will not be powerful enough to perform a blind survey, although we
examine the number counts in the standard radio bands. We consider the
detection-rate of molecular emission lines in two types of observing
campaigns:

\begin{enumerate}

\item Individual pointings  of  one hour  integrations searching  for
  $\geq5\sigma$   detections  of   lines  in   300\,km\,s$^{-1}$
  ($\Delta \nu = \nu/10^{3}$\,GHz)   bins
  (i.e.\ an integrated line detection where information about the line
  profile is discarded) across a set bandwidth. This strategy could be
  used for a controlled (i.e.\ flux limited) survey for a particular
  line species.

\item Individual pointings that  reach a line flux limit corresponding
  to  the  the  knee  of  the integrated  counts  where  $N(>F)\propto
  F^{-2}$, thus optimizing  the detection rate of sources  for a fixed
  observing time, again for detections in 300\,km\,s$^{-1}$ channels
  (see Blain et al.\ 2000 and Carilli \& Blain\ 2002). This strategy
  could be used for a simple `redshift search', where one aims to
  simply identify large numbers of galaxies at high-$z$, or identify
  unique classes of object that are unlikely to be found by any other
  means.

\end{enumerate}

We note that that may be other routes to blind redshift surveys; for
example, sensitive Fourier Transform Spectrometers (FTSs) that take
advantage of the large fields of view of future submm telecopes (e.g.\
CCAT potentially has a usable field-of-view of 1$^\circ$). These could
also be used for blind tomographic surveys of \cii in the submm
atmospheric windows (Figure\ 2) in a similar manner
to narrowband surveys of emission line galaxies in the OIR bands.  

\subsubsection{Atacama Large Millimeter Array}

ALMA covers the frequency range 80--900\,GHz, which will possibly be extended
down to 40\,GHz with the addition of Bands 1 and 2, which would allow coverage
of low-$J$ CO and HCN transitions beyond $z\sim 1$ (Figure\,2). At completion,
ALMA will consist of $\sim$50 12\,m single feed dishes (on baselines of up to
15\,km), and we consider this as our model of `full power' ALMA. The
field-of-view (as defined by the FWHM of the primary beam) across Bands\,3--10
is $\sim$0.01--0.8\,arcmin$^2$, scaling as $\nu^{-2}$ and the ALMA receivers
can observe up to 8\,GHz of instantaneous bandwidth (dual polarization), made
up of two 4\,GHz-wide side-bands, separated by a gap of 8\,GHz. In this work,
for simplicity, we treat the instantaneous bandwidth as a continuous block.

The typical observing frequency (chosen to be close to the peak atmospheric
transmission in that band), field-of-view and sensitivity (in
300\,km\,s$^{-1}$ channels) for Bands 3--10 is given in Table\ 2. For the
latter, we assume full-power ALMA and average weather conditions\footnote{for
the sensitivity estimates, we have used the latest version of the ALMA
sensitivity calculator:\
https://almascience.nrao.edu/call-for-proposals/sensitivity-calculator}. The
limiting (5$\sigma$) line fluxes for both the 1\,hour single-shot and optimal
integration modes are given and for each case we list the total surface
density of CO $J_{\rm up }\leq10$, HCN $J_{\rm up }\leq5$ and \cii line
emitters in an 8\,GHz band. The detection rates are given by the time required
to reach the corresponding 1$\sigma$ channel sensitivity, combined with the
field-of-view. Two versions of the counts are given, corresponding to the
`virial' and `super-virial' models for the dense gas phase (\S3.1). Both
models assume $\xi_{\rm SF}=0.25$, a conservative estimate for the star
formation `mode' and thus the contribution of the quiescent gas phase (\S3.2).

Figure\ 6 and 7 show the integral counts $N(>F_{\rm line})$, split into the
individal line species for the two models, clearly demonstrating the
progression toward high-$J$/low-$z$ lines with increasing frequency, and the
dominance of \cii in the number counts in ALMA Bands 4--10 (cut-off at lower
frequencies as the line moves to very high redshifts). Figures\,8 and 9 show
the relative distribution of line species that would be detected when
operating at the optimal flux limit, again highlighting the dominance of \cii
in Bands 4--10, corresponding to redshifts of $0.5\lesssim z \lesssim 10$. As
expected, HCN forms a negligible contribution to the line counts. Due to the
dominance of \cii the virial and super-virial models predict similar yields of
line emitters in the ALMA Bands, but there are subtle differences in the
molecular line yields, as shown in Figures 8 \& 9, and we again note that in
reality the detection rate of CO lines will be higher, due to the minimalist
approach we take. Similarly, other bright far-infrared lines of nitrogen and
oxygen will contribute (Coppin et al.\ 2012 in prep).

We re-iterate that these counts should be considered lower limits to the
expected yield in a blind survey. Nevertheless -- as is not surprising -- even
when operating at the optimal flux limit which should give the most efficient
detection rate for a fixed observing time, the small fields-of-view at all
ALMA frequencies imply significant observational investments would be required
to perform efficient blind molecular line surveys using ALMA alone. Clearly
the high-frequency Bands $\geq$7 are impractical for blind surveys, given the
precipitous decline in field-of-view and sensitivity beyond 300\,GHz.
Nevertheless, although challenging, blind (optimal) surveys in Band 4 and 5
could be useful for detecting very high-$z$ galaxies close to the epoch of
re-ionization via their \cii emission, and intermediate redshift CO emitters.

As an example of the potential reward of a practical observing campaign,
consider a 100\,hour survey at a {\it fixed frequency tuning} in ALMA Band 4.
Operating at the optimal depth limit, this survey would yield $\sim$100 \cii
emitters (which would be ULIRG/HLIRG-class galaxies at the corresponding line
flux limit) at $z\sim12$. This is probably optimistic; it goes without saying
that, at this high redshift, our model of the infrared luminosity density is
highly uncertain -- it is almost certainly overestimated because it simply
extrapolates the slow decline in space density and characteristic luminosity
in the number counts model. Indeed we currently know of {\it no} galaxies
beyond $z>10$, and so there are literally no constraints, other than what can
be gleaned from the shape of the far-infrared background (which the
B\'ethermin et al.\ model successfully re-produces). Therefore, the observed
abundance of \cii emitters close to the epoch of re-ionisation could be an
extremely valuable probe of the history of re-ionisation. As an example of the
impact of the form of the early evolution of the infrared luminosity density
on the detection rate, if we modify the evolution of the B\'ethermin et al.\
model such that infrared luminosity density at $z=10$ is an order of magnitude
lower ($\sim$$10^{7}L_\odot$\,Mpc$^{-3}$, cf.\ Figure\ 4), then the same
`optimal' blind survey would expect detections at a rate of one galaxy per
{\it ten} hours of observation in Band 4.

What are the blind detection prospects for a putative population of CO-dark,
but $L_{\rm [CII]}/L_{\rm IR}$-boosted metal-poor systems we discussed in
section \S3.3.1? We naively assume that the space density evolution of such
systems follows that of LBGs. For the parameterisation of the evolving
co-moving number density we use the latest estimates of Bouwens et al.\
(2011), who constrain the rest-frame ultra-violet luminosity function at
$z\sim7$ and $z\sim8$ via an application of the Lyman Break drop-out technique
in very deep {\it Hubble Space Telescope} infrared and optical imaging.
Modelling the LBG luminosity function as a Schechter function, Bouwens et al.\
(2011) fit linear evolutions of the characteristic luminosity $M^\star$,
density normalisation $\phi^\star$ and faint end slope $\alpha$ that maximize
the likelihood of re-producing the formal luminosity function fits at
$z\sim4$, $5$, $6$, $7$ and $8$. We take this model and assume $\log\,(L_{\rm
IR}/L_\odot)=\log\,(L_{\rm bol}/L_\odot)=11.67-0.58(M_{\rm UV,AB}+21)$
(Bouwens et al.\ 2009) to estimate the co-moving space density of \cii-bright
galaxies in the ALMA bands. Applying the same optimal survey strategy as
above, this model suggests that a blind survey in Bands 4--6 would detect
$\sim$2--7 galaxies per hour at redshifts of $z\sim8$--$12$. Again, we caution
that this estimate relies on extrapolation of the evolution of the LBG
luminosity function beyond current observational constraints, and the
high-{\it z} LBG luminosity function might not necessarily reflect that of a
population of metal-poor galaxies in the $L_{\rm [CII]}/L_{\rm IR} = 10^{-2}$
class. However, again, this highlights the discovery potential for blind
surveys at submm-to-mm wavelengths.

More realistically, ALMA will be routinely used in synergy with wide-field
sub-millimeter and radio continuum surveys (JCMT/SCUBA--2, LMT, CCAT, MeerKAT,
ASKAP, etc.), which will detect star-forming galaxies out to $z\sim10$ (taking
advantage of the negative {\it k}-correction in the sub-mm bands for instance)
that can be targeted for redshift identification. In this case our model can
be used to estimate the minimum line flux (and therefore exposure time
required) for redshift searches as a function of frequency (e.g.\ Figure\ 3).
In many cases the required flux limits can be achieved in a matter of minutes
with full-power ALMA, although a scan in frequency will be required for the
detection of several emission lines. Wide bandwidth submm direct detection
spectrographs (e.g.\ Z-Spectrometer) might be more practical for such a
targeted search, where several bright far-infrared and CO lines could be
detected simultaneously (Figure\ 2).

Finally, we note that an alternative application of this model is to predict
the number of serendipitous line detections in routine deep (e.g.\
$\sim$10\,hr) ALMA observations. Indeed, we can ask whether such data-cubes
could be `harvested' for line emitters in a semi-blind sense; exploiting the
more commonplace deep, pointed observations of some extragalactic source. In
Table\ 2 we list the number of serendipitous detections that would be expected
per field-of-view in a 10\,hour integration in each band. In Band\ 3, one
would expect $>$0.3--0.7 serendipitously detected sources per 10\,hour cube,
with detection rates naturally falling off with increasing frequency, and
declining sensitivity and field-of-view. Therefore, long after ALMA comes into
full operation, one could envision hunting for serendipitously detected
high-{\it z} line emitters in archival data.

\subsubsection{Square Kilometer Array and MeerKAT}

Phase 3 of the SKA will culminate in an array of 1250 15\,m single-feed
dishes; the current design plan is to include high-band coverage to
$\sim$30\,GHz, where the sensitivity will be $\sim$14\,$\mu$Jy\,$\surd$min\,
(in a 30\,MHz [300\,km\,s$^{-1}$] bin, dual polarization; S. Rawlings, 2011,
private communication). The field-of-view at 30\,GHz will be $\sim$7 square
arcminutes. However, SKA's high-frequency capability will only be available
towards the completion of the telescope, which is likely to be towards the end
of the next decade (and is therefore subject to design change); construction
will commence with the low frequency receivers.

One a shorter timescale, one of SKA's main pathfinders, the Extended Karoo
Array Telescope (MeerKAT) will have a high-band receiver covering 8--15\,GHz
(again, to be constructed in Phase 3 of the project. The other main SKA
pathfinder located in Western Australia -- ASKAP -- will not cover frequencies
beyond 2\,GHz and is therefore unsuitable for molecular surveys). MeerKAT will
consist of 64 13.5\,m single-feed dishes on baselines up to 20\,km. As shown
in Figure\,2, SKA/MeerKAT operating in the radio {\it K/Ka} bands will be
sensitive to \coi at $z\gtrsim3$, and thus will be vital for discovering the
molecular gas reservoirs fuelling galaxies in the very early Universe, close
to the epoch of re-ionization, $z\sim6$--$10$ (see\ Heywood et al.\ 2011).

In Table\ 2 we present the flux limits and number counts for the two observing
strategies, assuming 4\,GHz windows (MeerKAT and SKA are not likely to have
receivers capable of observing more than 4\,GHz of bandwidth). The power of
SKA's incredible sensitivity is clear here; in the {\it K} band even our
conservative estimates predict that $\sim$30--70 $z>3$ \coi--\coiii line
emitters could be detected every hour; at the optimal limit, the galaxies
would be in the ULIRG luminosity class (again CO emitters dominate the
detections). Figures\ 6--9 show the integral counts and line distribution,
which shows how SKA will open-up the $z>3$ Universe to low-{\it J} CO
exploration.

There is a clear synergy with ALMA here, since SKA will not be able to measure
the mid-to-high-{\it J} line emission in the galaxies it detects. In this
case, pointed observations of SKA detections with ALMA would be the natural
way to obtain robust redshifts and allow construction of the SLED of these
high-{\it z} galaxies. Armed with at least two lines for each galaxy, each
tracing different components of the gas reservoir, it would be possible to
perform another important survey -- the star formation `mode' of galaxies.
This is discussed in more detail in a follow-up work, Papadopoulos \& Geach
(2012, Paper\ II).

\section{Semi-blind redshift surveys}

Even when ALMA is at full capacity, blind molecular line searches will require
risky observational investments, albeit with the potential for rich reward.
Nevertheless, blind redshift surveys with ALMA are not out of the realms of
possibility, and the reward for such an endeavour could be vast. Blind,
high-redshift low-{\it J} CO surveys with SKA will be highly practical, given
the sensitivity of the instrument and the reasonable field-of-view.
Nevertheless, future redshift surveys in the submm-to-cm regime will likely
target large samples of galaxies pre-selected by their submm or radio
continuum emission; we call these semi-blind redshift surveys.

Our emergent model predicts the minimum CO, HCN and \cii line flux for a
galaxy with a given $L_{\rm IR}$ (Figure\ 3), and can therefore be used to
help design a semi-blind redshift survey by providing conservative estimates
for the exposure time required to detect a galaxy with some estimated $L_{\rm
IR}$. A semi-blind survey would have to perform a redshift search at the
position of each targeted galaxy, scanning in frequency until several emission
lines are detected (the CO ladder is spaced at intervals of
$\Delta\nu\sim$$115/(1+z)$\,GHz for example). Wide-band grating spectrometers
(e.g.\ Z-Spectrometer [Bradford et al.\ 2004] and ZEUS [Ferkinhoff et al.\
2010]) deployed in a multi-object capacity on ground-based telescopes will be
ideal for this purpose in the shorter wavelength submm windows (Figure\ 2).

Large-area sub-millimeter and radio continuum surveys during the next decade
will be able to supply thousands of targets for such a semi-blind survey. In
the short term, SCUBA--2 on the JCMT will be mapping $\sim$10\,deg$^2$ $S_{\rm
850}\sim$1.2\,mJy (1$\sigma$) depths, although this limit still corresponds to
rather luminous ($\sim$ULIRG-class) systems at $z\sim2$. Similarly {\it
Herschel} has recently mapped large areas of sky in the 250--500$\mu$m sub-mm
bands (Eales et al.\ 2010; Oliver et al.\ 2010), but these are relatively
shallow surveys suffering from significant confusion issues that, at redshifts
beyond $z\sim1$, are generally only tracing the most luminous galaxies, and
not probing into the $L^\star$ regime. Finally powerful radio galaxies at high
redshifts provide excellent beacons for such molecular line semi-blind
redshift surveys as they mark the centers of deep potential wells where
multiple gas-rich systems converge, forming the massive galaxy clusters found
in the present cosmic epoch (e.g.\ De Breuck et al. 2004; Miley \& De Breuck
2008).

In the near future, much larger single dish sub-millimeter telescopes such as
the LMT and CCAT will perform more sensitive, very wide area sub-mm surveys,
detecting the majority of the star formation rate budget out to $z\sim3$.
Offering similar promise is the imminent advent of all-sky sensitive radio
surveys during the next decade. For example, another SKA pathfinder, the
Australian SKA Pathfinder (ASKAP), will be performing a 1.3\,GHz radio
continuum survey called `EMU' (Evolutionary Map of the Universe). This will be
mapping the entire sky south of $\delta < +30^\circ$ to rms$\sim$10$\mu$Jy
sensitivity (Norris et al.\ 2011), detecting most of the star-forming galaxies
that exist out to $z\sim1$. One of the most critical aspects of semi-blind
surveys will be to properly understand the selection biases arising from a,
say, (sub)mm continuum flux limited or stellar mass selected sample, highlighting the need for a truly blind survey.

\section{Summary: fortune favours the brave}

We have presented a conservative model of the number counts of galaxies
detected in a blind molecular line survey in the sub-mm/mm/cm regime. Our
model calculates the `emergent' CO, HCN and \cii $\lambda$158$\mu$m emission
of star-forming galaxies, and is rooted in the latest models of star formation
feedback and empirical data on the HCN SLED (tracing the dense gas phase) in
local star-forming galaxies. The normalization of the emergent CO SLED is
given by the star formation rate, which in this case is taken to be the
infrared luminosity of a galaxy. Thus, our model describes the {\it minimum}
molecular line emission expected for star-forming galaxies based solely on the
luminosity of their actively star-forming reservoirs. This could be used to
design follow-up spectroscopic surveys for an unbiased $L_{\rm IR}$ limited
survey.

Coupled with an up-to-date model for the evolution of the infrared luminosity
density that successfully re-produces the observed number counts of galaxies
over a wide range of the infrared wavebands (B\'ethermin et al.\ 2011), we
make predictions of the lower limit of integrated number counts of
line-emitting galaxies across a range of observed frequencies and bandpasses
pertinent to the main facilities capable of performing a molecular redshift
survey (ALMA, SKA and its pathfinders). We consider ambitious blind redshift
surveys, working at the optimal flux limit set by the predicted knee in the
galaxy number counts, and discarding information about the shape of the
spectral line (i.e.\ binning to a spectral resolution of 1000, i.e.\
$\sim$300\,km\,s$^{-1}$). Such blind surveys can reveal insight into:

\medskip

{\it The epoch of re-ionization:}\ The sensitive ALMA bands could potentially
detect ULIRG-class \cii emitters close to the epoch of re-ionisation,
$z\gtrsim 10$, at a rate of up to one per hour (although this is highly
sensitive to the star formation history of the Universe at this early time).
Nevertheless, should such extreme systems exist at this epoch, a blind ALMA
survey would be capable of finding them, and their abundance would provide
valuable insight into the star formation and chemical history of the Universe
close to the era when the first stars ignited. In our minimal model, \cii
emitters dominate blind (optimal) surveys with ALMA, however mid-{\it J} CO
emitters would also be detected at lower rates, but with increasing yields for
deeper (but sub-optimal) surveys.

\medskip

{\it CO-dark galaxies:}\ We also examine the possibility of detecting \cii
luminous, but CO-dark gas reservoirs in metal-poor galaxies at high-{\it z}
with ALMA. Assuming such a population exists with a similar space density to
Lyman Break Galaxies, blind surveys with ALMA could detect systems at
$z\sim8$--$12$ with optimal rates of $\sim$2--7 per hour.

\medskip

{\it Efficient blind surveys of low-{\it J} CO emitters at $z\gtrsim3$:}\, The
SKA will represent a sea-change in the sensitivity of radio/cm-wave surveys,
with SKA Phase 3 (offering access to the radio {\it K} band) providing access
to low-{\it J} CO emission at $z>3$. We predict that an optimal redshift
survey could detect $\sim$30--70 ULIRG-class CO emitters per hour. While our
model is based on the abundance of star-forming galaxies, blind SKA surveys
could also detect outliers from the standard Schmidt-Kennicutt relation. In a
follow-up work, Paper\ II (Papadopoulos \& Geach 2012), we consider the
detectability of `pre-starburst' galaxies, representing a brief gas-rich phase
preceding the onset of an episode of intense star formation where the host
galaxy is extremely difficult to detect in any other waveband.

\medskip

\noindent The coming decade and the years beyond will be an exciting
time for extragalactic astronomy: we will routinely detect molecular
emission from high-redshift galaxies, breaking through the sensitivity
floor that has limited the majority of current studies to the most
luminous or fortuitously gravitationally lensed galaxies. This work
presents a simple, empirically-based model to aid in the design of
redshift surveys (both blind and semi-blind). Although we promote
ambitious observations, with -- arguably -- speculative results,  we
are motivated by the rich spoils: totally new and, in some cases, unique
insights into the physics of galaxy formation that could be the
reward for such efforts.

\section*{Acknowledgements}

We thank the referee for suggestions that improved the clarity of this paper.
J.E.G.\ is supported by a Banting Postdoctoral Fellowship administered by the
Natural Sciences and Engineering Research Council of Canada. The project was
funded also by the John S.\ Latsis Benefit Foundation. The sole responsibility
for the content lies with the authors. P.P.P.\ would like to thank the
Director of the Argelander Institute of Astronomy Frank Bertoldi, the
Rectorate of the University of Bonn, and the Dean U.-G. Meissner, for their
`Hausverbot' initiative that was a catalyst for finishing this work ahead of
schedule. We thank Matthieu B\'ethermin for assistance with the model of the
evolution of infrared luminosity density, and Ian Smail for valuable
discussions and suggestions for improvement. Finally, we acknowledge helpful
information on the SKA design provided by Steve Rawlings, who sadly passed
away during the completion of this work.

\label{lastpage}


\begin{thebibliography}{100}
	

\bibitem[Author et al. (2012)]{author}{Aalto S., Booth R. S., Black J. M., \& Johansson L. E. B. 1995, A\&A, 300, 369}

\bibitem[Author et al. (2012)]{author}{ Allen R. J., Le Bourlot J., Lequeux J., Pineau des Forets G., \& Roueff E. 1995, ApJ, 444, 157}

\bibitem[Author et al. (2012)]{author}{Andrews B. H., \&  Thompson T. A. 2011, ApJ, 727, 97}


\bibitem[Author et al. (2012)]{author}{Baker A., Tacconi L. J., Genzel R., Lehnert M. D., \& Lutz D. 2004, ApJ, 604, 125}

\bibitem[Author et al. (2010)]{author}{B\'ethermin, M., Dole, H.,
Lagache, G., Le Borgne, D., Penin, A., 2011, A\&A, 529, 4}

\bibitem[Author et al. (2012)]{author}{Blain A. W., Frayer D. T., Bock
J. J., Scoville N. Z., 2000, MNRAS, 313, 559}

\bibitem[Author et al. (2012)]{author}{Bolatto A. D, Jackson J. M., \& Ingalls J. G. 1999, ApJ, 513, 275}

\bibitem[Authoer et al. (2012)]{author}{Bouwens R. J., et al., 2009,
ApJ, 705, 936}


\bibitem[Authoer et al. (2012)]{author}{Bouwens R. J., et al., 2011,
ApJ, 737, 90}

\bibitem[Author et al. (2012)]{author}{Bradford C., et al.\ 2004,
Millimeter and Submillimeter Detectors for Astronomy II, Eds:
Zmuidzinas, J., Holland, W., Withington, S., Proceedings of the SPIE,
Vol 5498, pp. 257}



\bibitem[Author et al. (2012)]{author}{Braine J. \& Combes F. 1992, A\&A, 264, 433}

\bibitem[Author et al. (2012)]{author}{Brauher J. R., Dale, D. A.,
Helou, G., 2008, ApJS, 178, 280}

\bibitem[Author et al. (2012)]{author}{Brown R. L. \& Vanden Bout P. A. 1991, AJ, 102, 1956}

\bibitem[Author et al. (2012)]{author}{Bryant P M., \& Scoville N. Z. 1996, 
ApJ, 457, 678}

\bibitem[Author et al. (2012)]{author}{Carilli C. L., Blain A. W.,
2002, ApJ, 569, 605 }


\bibitem[Author et al. (2012)]{author}{Combes F., Maoli R., Omont A.,
1999, A\&A, 345, 369}

\bibitem[Author et al. (2010)]{author}{Coppin et al.\ 2007, ApJ, 665,
936}


\bibitem[Author et al. (2012)]{author}{Daddi E., Bournaud F., Walter F. et al. 2010, ApJ, 713, 686}

\bibitem[Author et al. (2012)]{author}{Danielson A. L. R. Swinbank A. M., Smail I. et al. 2011, MNRAS, 410, 1687}

\bibitem[Author et al. (2012)]{author}{Dannerbauer H., Daddi E., Riechers D. A., Walter F., Carilli C. L., Dickinson M., Elbaz D., \& Morrison, G. E. 2009, ApJ, 698, L178}


\bibitem[Author et al. (2012)]{author}{De Breuck, C., et al. 2004, A\&A, 424, 1}

\bibitem[Author et al. (2012)]{author}{De Breuck, C. Downes D., Neri R., van Breugel W., Reuland M., Omont A., \& Ivison R. 2005, A\&A, 430, L1}



\bibitem[Author et al. (2012)]{author}{Eales, S., et al.\ 2010, PASP,
122, 499}

\bibitem[Author et al. (2012)]{author}{Gao Y., \& Solomon P. M. 2004, ApJ, 606, 271}

\bibitem[Author et al. (2012)]{author}{Greve T. R., Bertoldi F., Smail I., et 
al. 2005, MNRAS, 359, 1165}

\bibitem[Author et al. (2012)]{author}{Greve, T. R., Papadopoulos, P. P., Gao, Y., Radford, S. J. E., 2009, ApJ, 692, 1432}

\bibitem[Author et al. (2012)]{author}{Ferkinhoff, C., Nikola, T.,
Parshley, S. C., Stacey, G. J., Irwin, K. D., Cho, H.-M., Halpern, M.,
2010,
Millimeter, Submillimeter and Far-Infrared Detectors and
Instrumentation for Astronomy V., Eds: Holland, W. S., Zmuidzinas, J.,
Proceedings of the SPIE, Vol 7741, pp 77410Y-77410Y-14}

\bibitem[Author et al. (2012)]{author}{Feruglio C., Maiolino R., Piconcelli E., Menci N., Aussel H., Lamastra A., \&  Fiore F. 2010, A\&A, 518, L155}

\bibitem[Author et al. (2012)]{author}{ Fixsen D. J.,  Bennett C. L., \& Mather J. C. 1999, ApJ, 526, 207}

\bibitem[Author et al. (2012)]{author}{Frayer D. T., Ivison R. J., Scoville N. Z., Yun M., Evans A. S., Smail I., Blain A. W., \& Kneib J.-P. 1998,
 ApJ, 506, L7}

\bibitem[Author et al. (2012)]{author}{Frayer D. T., Ivison R. J., Scoville N. Z.,  Evans A. S., Yun M. S.,  Smail I., Barger A. J., Blain A. W.,
  \&  Kneib J.-P. 1999, ApJ, 514, L13}


\bibitem[Author et al. (2012)]{author}{Heywood, I., et al.\ 2011,
Astronomy with megastructures: Joint science with the E-ELT and SKA,
Eds: Hook, I., Rigopoulou, D., Rawlings, S. \& Karastergiou, A., arXiv1103.0862}


\bibitem[Author et al. (2012)]{author}{Israel F. P., Tilanus R. P. J., \& Baas F. 1998, A\&A, 339, 398}	


\bibitem[Author et al. (2012)]{author}{Jackson J. M., Paglione T. A. D., Carlstrom J. E., \& Nguyen-Q-Rieu 1995, ApJ, 438, 695}

\bibitem[Author et al. (2012)]{author}{Juneau S., Narayanan D., Moustakas J., Shirley Y. L., Bussmann R. S., Kennicutt R. C. Jr., 
\& Vanden Bout P. A. 2009, ApJ, 707, 1217}


\bibitem[Author et al. (2012)]{author}{Kennicutt, R. C.\ Jr., 1998, ApJ, 498,
541}


\bibitem[Author et al. (2012)]{author}{Krips, M., Neri, R., Garc\'ia-Burillo, S., Mart\'in, S., Combes, F., Graci\'a-Carpio, J., Eckart, A., 2008, APJ, 677, 262}

\bibitem[Author et al. (2012)]{author}{ Krumholz M., \& McKee C. F. 2005, ApJ, 630, 250}

\bibitem[Author et al. (2012)]{author}{Lagache G., Dole H., Puget
J.-L., 2003, MNRAS, 338, 555}


\bibitem[Author et al. (2012)]{author}{ Loinard L., Allen R. J., \&  Lequeux J. 1995, A\&A, 301, 68 }

\bibitem[Author et al. (2012)]{author}{ Loinard L., Allen R. J., \&  Lequeux J. 1996, A\&A, 310, 93}

\bibitem[Author et al. (2012)]{author}{ Loinard L., Allen R. J. 1998, ApJ, 499, 227}

\bibitem[Author et al. (2012)]{authoer}{Luhman M. L., Satyapal S.,
Fischer J., Wolfire M. G., Sturm E., Dudley C. C., Lutz D., Genzel R.,
2003, ApJ, 594, 758}

\bibitem[Author et al. (2012)]{authoer}{Lupu, R. E. et al., 2011, 2010arXiv1009.5983}

\bibitem[Author et al. (2012)]{author}{Madden S. C., Poglitsch A.,
Geis N., Stacey G. J., \& Townes C. H., 1997, ApJ, 483, 200}

\bibitem[Author et al. (2012)]{authoer}{Malhotra S., et al., 2001,
ApJ, 561, 766}

\bibitem[Author et al. (2012)]{author}{Maiolino R., et al., 2009, A\&A, 500, L1}

\bibitem[Author et al. (2012)]{author}{Mao R. Q., Schulz A., Henkel C., Mauersberger R., Muders D., \& Dinh-V-Trung 2011, ApJ, 724, 1336}

\bibitem[Author et al. (2012)]{author}{ Mauersberger R., Henkel C. Walsh W., \& Schulz A. 1999, A\&A, 341, 256}


\bibitem[Author et al. (2012)]{author}{Meijerink R., \& Spaans M. 2005, A\&A, 436, 397}

\bibitem[Author et al. (2012)]{author}{Miley G. \& De Breuck C. 2008,
	The Astronomy and Astrophysics Review, Volume 15, Issue 2, pp.\ 67}
	

\bibitem[Author et al. (2012)]{author}{Narayanan D., Krumholz M.,
Ostriker E. C., \& Hernquist L. 2011, MNRAS, 418, 664}

\bibitem[Author et al. (2012)]{author}{Nieten C., Dumke M., Beck R., \& Wielebinski R. 1999, A\&A, 347, L5}

\bibitem[Author et al. (2012)]{author}{Nguyen-Q-Rieu, Nakai N., \& Jackson J. M. 1989, A\&A, 220, 57}

\bibitem[Author et al. (2012)]{author}{Norris, R. P., et al., 2011,
PASA, 28, 215}


\bibitem[Author et al. (2012)]{author}{Oliver, S. J., 2010, A\&A, 518,
21}

\bibitem[Author et al. (2012)]{author}{Pak S. et al. 1998, ApJ, 498,
735}

\bibitem[Author et al. (2012)]{author}{Papadopoulos, Thi, \& Viti,
2002, ApJ, 579, 270}

\bibitem[Author et al. (2012)]{author}{Papadopoulos P. P., \& Greve T. R. 2004, ApJ, 615, L29}

\bibitem[Author et al. (2012)]{author}{Papadopoulos P. P., Isaak K. G., \& van der Werf P. P. 2007, ApJ, 668, 815}

\bibitem[Author et al. (2012)]{author}{Papadopoulos P. P. 2010, ApJ, 720, 226}

\bibitem[Author et al. (2012)]{author}{Papadopoulos P. P. \&  Pelupessy F. I. 2010, ApJ, 717, 1037}

\bibitem[Author et al. (2012)]{author}{Papadopoulos P. P. \& Seaquist E. R. 	1999, ApJ, 516, 114}

\bibitem[Author et al. (2012)]{author}{Papadopoulos, P.\ P. et al. 2012 ApJ, 751, 10}

\bibitem[Author et al. (2012)]{author}{Paglione T. A. D.,  Tosaki T., \& Jackson J. M. 1995, 454, L117}

\bibitem[Author et al. (2012)]{author}{Paglione T. A. D., Jackson J. M., \& Ishizuki S. 1997, ApJ, 484, 656}

\bibitem[Author et al. (2012)]{author}{Pelupessy F. I., Papadopoulos, Padeli P.; van der Werf, P., 2006, ApJ, 645, 1024}

\bibitem[Author et al. (2012)]{author}{Penzias A. A., Jefferts K. B., \& Wilson R. W. 1971, ApJ, 165, 53}

\bibitem[Author et al. (2012)]{author}{Penzias A. A., Solomon P. M., Jefferts K. B., \& Wilson R. W. 1972, ApJ, 174, L43}

\bibitem[Author et al. (2012)]{author}{Rodighiero G., et al., 2010,
A\&A, 515, 8}

\bibitem[Author et al. (2012)]{author}{Rickard L. J., Palmer P., Morris M., Zuckerman B. \& Turner B. E. 1975, ApJ, 199, L75}

\bibitem[Author et al. (2012)]{author}{Sakamoto  K.,  Aalto S.,  Wilner, D. J. et al. 2009, ApJ, 700, 104}


\bibitem[Author et al. (2012)]{author}{Scoville N. Z. 2004, in {\it The Neutral ISM in Starburst Galaxies}, Astronomical  Society of the Pacific Conference Series, Vol 320, pg., 253}

\bibitem[Author et al. (2012)]{author}{Solomon P. M., Downes D., Radford S. J. E., \& Barrett J. W. 1997, ApJ, 478, 144}

\bibitem[Author et al. (2012)]{author}{Solomon P. M., Downes D., \& Radford S. J. E. 1992a, ApJ, 387, L55}

\bibitem[Author et al. (2012)]{author}{Solomon P. M., Radford S. J. E., \& Downes D. 1992b, Nature, 356, 318}

\bibitem[Author et al. (2012)]{author}{Solomon P. M., Vanden Bout, P.\ A.
2005, ARA\&A, 43, 677}

\bibitem[Author et al. (2012)]{author}{Stacey G. J., et al.\ 2010,
ApJ, 724, 957}

\bibitem[Author et al. (2012)]{author}{Swinbank A.\ M., et al. 2011,
ApJ, 742, 11}


\bibitem[Author et al. (2012)]{author}{Thompson T. A., Quataert E., \& Murray N. 2005, ApJ, 630, 167}

\bibitem[Author et al. (2012)]{author}{Thompson T. A. 2009, in {\it The Starburst-AGN connection},  Astronomical Society of the Pacific Conference Series, Vol 408, pg. 128}

\bibitem[Author et al. (2012)]{author}{Walter F., Bertoldi F., Carilli C. et al. 2003, Nature, 424, 406}

\bibitem[Author et al. (2012)]{author}{ Walter F., Carilli C., Bertoldi F., Menten K., Cox P., Lo K. Y., Fan X., \& Strauss M. A. 2004, ApJ, 615, L17}

\bibitem[Author et al. (2012)]{author}{Walter F., \& Carilli C. 2008, Ap\&SS, 313, 313}

\bibitem[Author et al. (2012)]{author}{Wang J., Zhiuy Z., \& Yong S. 2011, MNRAS, 416, L21}

\bibitem[Author et al. (2012)]{author}{Weiss A., Downes D., Walter F., \& Henkel C. 2007, ASP Conference Series, Vol 375, pg. 25}

\bibitem[Author et al. (2012)]{author}{Wilson R. W., Jefferts K. B. \& Penzias A. A. 1970, ApJ, 161, L43}

\bibitem[Author et al. (2012)]{author}{Wilson C. D. 1997, ApJ, 487, L49}

\bibitem[Author et al. (2012)]{author}{Wu J., Evans N. J. II, Gao Y., Solomon P. M., Shirley Y. L., \& Vanden Bout P. A. 2005, ApJ, 635, L173}

\bibitem[Author et al. (2012)]{author}{Yao L., Seaquist E. R. Kuno N., Dunne L. 2003, ApJ, 588, 771}

\bibitem[Author et al. (2012)]{author}{ Young J. S. \& Scoville N. Z. 1991, ARA\&A 29, 581}

\end{thebibliography}
\end{document}